\documentclass[useAMS]{mn2e}

\usepackage{amssymb,amsmath,psfig,times}
\def\gsim{ \lower .75ex \hbox{$\sim$} \llap{\raise .27ex \hbox{$>$}} }
\def\lsim{ \lower .75ex\hbox{$\sim$} \llap{\raise .27ex \hbox{$<$}} }

\title[Testing a new view of Gamma Ray Burst Afterglows]
{Testing a new view of Gamma Ray Burst Afterglows}  
\author[M. Nardini, G. Ghisellini, G. Ghirlanda, A. Celotti]
{M. Nardini$^1$\thanks{Email:
nardini@sissa.it}, G. Ghisellini$^2$,  G. Ghirlanda$^2$ and
A. Celotti$^1$. \\
$^1$SISSA -- Via Beirut 2-4, 34151 Trieste, Italy\\
$^2$INAF -- Osservatorio Astronomico di Brera,  Via Bianchi 46, 23806 Merate, Italy
}

\begin{document}  

\maketitle

\begin{abstract}
The optical and X--ray light--curves of long Gamma Ray Bursts (GRBs) often 
show a complex evolution and in most cases do not track each other. 
This behaviour can not be easily explained by the 
simplest standard afterglow models. 
A possible interpretation is to consider the observed optical and 
X--ray light--curves as the sum of two separate components.  
This scenario requires the presence of a spectral break between these 
bands.  One of the aims of this work is to test whether such a break is present within
the observed {\it Swift} XRT energy  range.
We analyse the X--ray afterglow spectra of a sample of 33 long GRBs 
with known redshift, good optical photometry and published estimate 
of the host galaxy dust absorption $A_{\rm V}^{\rm host}$. 
We find that  indeed  in 7 bright events  
a broken power--law provides 
a  fit to the data that is better than a single power--law model. 
For 8 events, instead, the X--ray spectrum is better fitted by 
a single power--law. 
We discuss the role of these breaks  in connection to the relation between 
the host hydrogen column density 
$N_{\rm H}^{\rm host}$ and $A_{\rm V}^{\rm host}$ and  check 
the consistency of the X--ray spectral breaks with the optical bands photometry.
We analyse the optical to X--ray spectral energy distributions at different 
times and find again consistency with two components interpretation.  

\end{abstract}


\section{Introduction}

The fast re--pointing capabilities of the {\it Swift} satellite 
(Gehrels et al. 2004) allowed to  reveal the early time afterglow 
behaviour of Gamma--Ray Bursts (GRBs) and its unforeseen complexity.
 Several interpretation have been proposed 
to account for  the unexpected light--curves evolution. 
This is often characterised by an early time steep flux decay starting 
after the end of the gamma-ray prompt emission, followed by a shallower 
 and a subsequent steeper decay, the latter 
corresponding to the typical afterglow observed in the pre--{\it Swift} era  (see e.g. Nousek et al. 2006, Zhang et al. 2006).

Great efforts 
have been made to explain the  origin  of the shallow
decay phase which can last from some hundreds  up to hundred of 
thousand seconds  as it cannot be explained in the frame of the simplest 
``standard" forward shock fireball models. 
In  Ghisellini et al. (2009) (hereafter G09), we presented a summary 
of the proposed  interpretations (see also Zhang 2007 for  a review). 

In the last years, 
the increasing 
number of well sampled light--curves allowed to  examine simultaneous  optical and 
X--rays afterglow light--curves  of several 
long GRBs (e.g. Curran et al. 2009,  De Pasquale et al. 2009).
In G09 we analysed the broad band optical to X--ray rest frame 
temporal behaviour of a sample of 33 GRBs observed by { \it Swift} XRT with 
known redshift, published host galaxy dust absorption estimate 
and good  quality optical follow up. 
In some cases the optical and 
X--ray temporal evolution are very different.   
We proposed that the light curves behaviour is due 
to the sum of two separate components. 
The first one is assumed to  originate from a standard forward (external) 
shock, as described by Panaitescu \& Kumar (2000). The 
second component is treated in a purely phenomenological way with 
the aim of minimising the number of free parameters. 
 A possible physical origin for it 
can be provided within the so called ``late prompt" 
scenario described by Ghisellini et al. (2007). 
In G09 we found that this two component modelling is able to well reproduce 
all  of the optical and X--ray light curves of the GRBs of the sample 
(once the early steep decay phase and the flaring 
activity that sometimes appears in the X--ray light--curves are excluded). 

In order to  test the consistency of the two components interpretation and make a first step towards a more 
physical scenario, it is important to  verify  whether the observed X--ray 
spectra and the optical to X--rays spectral energy distributions (SEDs) 
are in agreement with what predicted by the light--curves modelling. 

In a scenario where the optical and the X--ray emission are due to different 
 processes, the component accounting for the optical spectrum has to break 
in order not to dominate also in the X--ray band. 
Conversely   a
break to a harder spectral index is also required towards the soft end 
of the X-ray component, not to interfere with the optical  emission.
However, a spectral break (e.g. the cooling break frequency of the 
synchrotron emission mechanism) between the optical and X--ray  bands is 
sometimes expected also in the standard afterglow scenario (see  e.g. Figs. 10 
and 11  in Nardini et al. 2006).
In both scenarios
the spectral break can be at frequencies within the  
observed XRT 0.3--10 keV energy range. 
 If this is the case, spectral fits provide not only the break frequency but 
also the spectral slope below it.
This additional information makes  
these events the best   candidates to test 
the two component light--curve 
modelling from a spectral point of view. 
If the  optical and X--ray light--curves are dominated by the same component, the  
observed optical fluxes must be consistent with the extrapolation of  
the X--ray low energy spectrum. 
If the  light--curves
are  instead dominated by different components, the X--ray  spectrum extrapolation  
should  not significantly contribute to the observed optical flux  (see \S \ref{sed} for 
a more detailed discussion).

In this work we analyse the XRT spectra of the GRBs in the G09 sample  to check for the 
presence of such a break.  
In order to test whether the  X--ray spectral break is
consistent with what seen (simultaneously) in the optical,
in \S \ref{sed} we  examine the optical to X--ray SEDs
 sampled at different times along the light--curves. 
 Such a combined analysis of  broad band light--curves and 
optical to X--ray SEDs represents  a crucial consistency check for  
our proposed interpretation.

 As discussed in section \S \ref{nhav} an interesting outcome of the spectral analysis is related to the apparent  
``discrepancy"
between the  amount  of the X--ray 
absorption (as measured
by the  hydrogen column density $N_{\rm H}^{\rm host}$) and the optical extinction
$A_{\rm V}^{\rm host}$ in the host frame.
The  value of $N_{\rm H}^{\rm host}$  inferred from fitting the X--ray spectra with a 
single power--law is often at odds  (for standard gas-to-dust conversions) with
the relatively small $A_{\rm V}^{\rm host}$ evaluated
through the analysis of the optical SEDs
(see  e.g. Galama \& Wijers 2001; Stratta et al. 2004; Kann, Klose \& Zeh 2006, Schady  et al. 2007).
If the intrinsic X--ray spectrum  can be well modelled by a broken power--law,
then the required $N_{\rm H}^{\rm host}$  is smaller than what required by a single power--law fit),
ameliorating the $N_{\rm H}^{\rm host}$--$A_{\rm V}^{\rm host}$ disagreement.

\section{The sample}

 The sample comprises the 33 long GRBs considered G09, whose selection criteria were: 
the knowledge of the GRB redshift, a good photometric coverage, 
{\it Swift} XRT observations and a published estimate of the 
host galaxy dust  absorption  $A_{\rm V}^{\rm host}$.
When different values of $A_{\rm V}^{\rm host}$ 
are reported in the literature we chose
the estimate derived from a direct analysis of the optical spectral
energy distribution rather then that obtained by a combined
analysis of the optical to X--ray SEDs.  If 
only the latter is available we discuss the effects of 
possible alternative solutions, through a  direct analysis of the  SEDs (see \S \ref{sed}).  

\section{XRT data reduction and spectral analysis}
\label{xrtspectra}

We analysed the XRT data of  the events in the sample with the {\it Swift} 
software package {\it v2.9} distributed with 
HEASOFT ({\it v6.6}).  The XRT data were reprocessed  with the XRTPIPELINE 
tool\footnote{Part of the XRT software,
  distributed with HEASOFT package:  {\tt http://heasarc.gsfc.nasa.gov/heasoft/}}.  
The spectra were extracted in both WT and PC mode with the standard grade, applying, when required, 
the correction for pileup (Moretti et al. 2005, Romano et al. 2006; 
Vaughan et al. 2006).  The extraction was 
in boxes (WT mode) or circular regions (PC mode) of typical widths as discussed 
in Evans et al. (2009). 
Background spectra were extracted in same-sized regions far from the source. 
For all  of the spectra we created Ancillary Response Files with the \texttt{xrtmkarf} 
tool and used the calibration database updated to December 2008. 
 The spectra were re-binned in order to 
have a minimum of 20 counts per energy  bin (15 for the faintest events) and energy 
channels below 0.3 keV and above 10 keV were excluded from the  analysis. 
 The XSPEC({\it v11.3.2}) software was utilised for the analysis. 
For bursts with particularly bright X-ray emission we also performed a time resolved spectral 
analysis in order to check for the possible spectral  evolution.
Since we are not
considering XRT data that are simultaneous to the BAT $\gamma$--ray
detection, the steep early time phase and the flaring activity are not
considered. 

\subsection{Single absorbed power--law model}
\label{abp}

Following the conventional analysis of X-ray GRB spectra we  fitted
all the spectra  with a model composed by a power--law with
two absorption components at low energies, \texttt{wabs} and \texttt{zwabs}. 
The first one corresponds to  Galactic absorption and its column density 
$N_{\rm H}^{\rm gal}$ is fixed to the Galactic value (from Kalberla et al.  2005).
The second absorption is due to the material located at the redshift of the source and its column density
$N_{\rm H}^{\rm host}$  is let free to vary. 
The 90\% confidence intervals on the best fit parameters are obtained with the 
\texttt{error} command in XSPEC.
All the spectra returned a good fit with such a model, with  
$\chi^2/{\rm dof}$  close to unity.  The best fit parameters are in a good 
agreement both with the results of the automatic XRT data analysis tool available on line\footnote{{\tt http://www.swift.ac.uk/xrt\_spectra/}}
developed by Evans et al. (2008, 2009)  
and with the values  reported in the literature (summarised also in Tab. 1 of G09). 
The results of the  fits are reported in Tab. \ref{nh_av_pow}. 
  
\begin{table*}
\begin{center}
\begin{tabular}{llllllll}
\hline
\hline
GRB &$z$&$t_{\rm start}-t_{\rm end}$&$\beta_{\rm X}$ &$N_{\rm H}^{\rm host}$  &$\chi^2_{\rm R}$ (dof)&$A_{\rm V}^{\rm host}$&Ref \\ 
        &      &   s after trigger            &
&  $10^{21}$ cm$^{-2}$   &  &magnitudes                        &       \\ 
\hline
\hline
050318  &1.44   &3.3$\times 10^3$-6.3$\times 10^4$ &1.1$\pm$0.1   &0.5$\pm$0.4  &0.89 (80)  &0.68$\pm$0.36   &Ber05a, Sti05  \\    
050319  &3.24   &5.0$\times 10^3$-1.1$\times 10^5$ &1.06$\pm$0.12 &4.$\pm$4.    &0.76 (46)  &0.11            &Fyn05a, Kan09 \\
050401  &2.8992 &1.3$\times 10^2$-8.5$\times 10^3$ &0.88$\pm$0.04 &15.6$\pm$1.9 &1.056 (273)&0.62$\pm$0.06   &Fyn05b, Wat06 \\ 
050408  &1.2357 &2.6$\times 10^3$-7.1$\times 10^4$ &1.15$\pm$0.16 &12.2$\pm$2.8 &1.36 (37)  &0.73$\pm$0.18   &Ber05b, dUP07\\
050416A &0.653  &3.5$\times 10^2$-1.5$\times 10^5$ &1.01$\pm$0.11 &5.8$\pm$1.1  &0.88(74)   &0.19$\pm$0.11   &Cen05, Hol07\\ 
050525A &0.606  &5.9$\times 10^3$-7.4$\times 10^4$ &1.1$\pm$0.17  &2.1$\pm$1.1  &0.86 (32)  &0.32$\pm$0.2    &Fol05, Kan09 \\ 
050730  &3.967  &1.5$\times 10^4$-1.4$\times 10^5$ &0.62$\pm$0.08 &4.8$\pm$4.8  &1.24 (95)& 0.01$\pm$0.005 &Che05, Sta05 \\ 
050801  &1.56   &6.5$\times 10^2$-5.2$\times 10^4$ &0.84$\pm$0.20 &0$\pm$0.07   &0.66 (14)  &0               &DeP07, Kan09  \\
050802  &1.71   &4.8$\times 10^2$-9.3$\times 10^4$ &0.82$\pm$0.06 &1.8$\pm$1.0  &1.055 (159)&0.55$\pm$0.1    &Fyn05c Sch07   \\
050820A &2.612  &4.7$\times 10^3$-5.9$\times 10^4$ &0.99$\pm$0.06 &3.3$\pm$2.2  &0.98 (143) &0.065$\pm$0.008 &Pro05, Kan09  \\ 
050824  &0.83   &6.6$\times 10^3$-1.0$\times 10^5$ &0.87$\pm$0.18 &0.6$\pm$0.6  &0.95 (32)  &0.14$\pm$0.13   &Fyn05d, Kan09  \\ 
050922C &3.221  &1.1$\times 10^2$-4.5$\times 10^2$ &1.02$\pm$0.07 &3.6$\pm$2.2  &1.00 (115) &0               &Jak06, Kan09  \\
051111  &1.55   &5.6$\times 10^3$-5.3$\times 10^4$ &1.21$\pm$0.19 &6.1$\pm$3.0  &0.80 (36)  &0.39$\pm$0.11   &Hil05, Sch07  \\ 
060124  &2.296 	&3.4$\times 10^4$-1.2$\times 10^5$ &1.02$\pm$0.08 &7.6$\pm$2.5  &0.81 (107) &0               &Cen06b, Mis07   \\ 
060206  &4.045  &5.1$\times 10^3$-3.5$\times 10^4$ &1.29$\pm$0.15 &15.3$\pm$9.5 &0.99 (87)  &0$\pm$0.02      &Fyn06, Kan09  \\ 
060210  &3.91   &3.8$\times 10^3$-5.8$\times 10^4$ &1.10$\pm$0.06 &17.5$\pm$5.0 &1.017 (185)&1.1$\pm$0.2     &Cuc06, Cur07 \\ 
060418  &1.489  &2.6$\times 10^2$-6.7$\times 10^2$ &0.87$\pm$0.09 &4.2$\pm$1.7  &0.86 (91)  &0.25$\pm$0.22   &Pro06, Ell06 \\
060512  &0.4428 &3.7$\times 10^3$-2.3$\times 10^5$ &0.97$\pm$0.18 &0.2$\pm$0.2  &1.39 (17)  &0.44$\pm$0.05   &Blo06, Sch07\\
060526  &3.221  &7.4$\times 10^2$-7.6$\times 10^3$ &0.95$\pm$0.13 &6.$\pm$6.    &0.59 (31)  &0.04$\pm$0.04   &Ber06, Th\"o08  \\ 
060614  &0.125  &4.4$\times 10^3$-2.8$\times 10^4$ &0.79$\pm$0.09 &0.3$\pm$0.3  &0.98 (66)  &0.05$\pm$0.02   &Pri06, Man07   \\
060729  &0.54   &1.7$\times 10^4$-1.8$\times 10^5$ &1.05$\pm$0.02 &0.9$\pm$0.2  &1.01 (290) &0.              &Th\"o06, Gru07    \\
060904B &0.703  &1.0$\times 10^3$-4.1$\times 10^4$ &1.10$\pm$0.12 &2.7$\pm$1.2  &0.86 (40)  &0.44$\pm$0.05   &Fug06, Kan09    \\ 
060908  &2.43   &1.5$\times 10^2$-1.9$\times 10^3$ &0.84$\pm$0.11 &2.$\pm$2.    &1.09 (60)  &0.055$\pm$0.033 &Rol06, Kan09  \\ 
060927  &5.47   &1.0$\times 10^2$-6.1$\times 10^3$ &0.9$\pm$0.2   &0.5$\pm$0.5  &0.75 (15) &0.33$\pm$0.18   &Fyn06b, RuV07   \\ 
061007  &1.26   &2.0$\times 10^2$-2.1$\times 10^3$ &0.91$\pm$0.02 &5.6$\pm$0.3  &1.054 (480)&0.54$\pm$0.32   &Osi06,  Kan09 \\ 
061121  &1.314  &2.0$\times 10^2$-1.8$\times 10^4$ &1.01$\pm$0.08 &7.3$\pm$1.3  &0.88 (121) &0.72$\pm$0.06   &Blo06, Pag07  \\ 
061126  &1.1588 &1.8$\times 10^3$-1.5$\times 10^4$ &0.81$\pm$0.11 &5.6$\pm$1.2  &1.08 (143) &0               &Per08a, Per08a    \\ 
070110  &2.352  &4.0$\times 10^3$-4.5$\times 10^4$ &1.12$\pm$0.07 &2.6$\pm$1.5  &0.875 (129)&0.08            & Jau07, Tro07\\ 
070125  &1.547  &4.7$\times 10^4$-1.3$\times 10^5$ &0.97$\pm$0.2  &1.7$\pm$1.7  &0.88 (21)  &0.11$\pm$0.04   &Fox07, Kan09   \\ 
071003  &1.604  &2.2$\times 10^4$-4.2$\times 10^4$ &1.95$\pm$0.12 &0.7$\pm$0.7  &1.20 (47)  &0.209$\pm$0.08  &Per07, Per08b  \\ 
071010A &0.98   &3.4$\times 10^4$-9.1$\times 10^4$ &1.43$\pm$0.5  &13.5$\pm$7.0 &0.52 (11)  &0.615$\pm$0.15  &Pro07, Cov08a    \\ 
080310  &2.42   &1.7$\times 10^4$-5.2$\times 10^4$ &0.85$\pm$0.1  &3.0$\pm$3.0  &1.11 (36)  &0.1$\pm$0.05    &Pro08, PeB08\\ 
080319B &0.937  &5.6$\times 10^2$-1.7$\times 10^3$ &0.80$\pm$0.01 &1.6$\pm$0.1  &1.35 (610) &0.07$\pm$0.06   &Vre08, Blo09   \\ 
\hline
\hline
\end{tabular}
\caption{Results of the single power--law fitting.  For each GRB we report:
  the redshift, the time interval in which  the
  spectrum was extracted, the unabsorbed spectral index $\beta_{\rm X}$,  the hydrogen
  column density at the host $N_{\rm H}^{\rm host}$, the reduced $\chi^2$
  and number of degrees of freedom, the host galaxy visual extinction
  $A_{\rm V}^{\rm host}$ taken from the literature, and the references for
  redshift and $A_{\rm V}^{\rm host}$.  
References: 
Ber05a: Berger et al. (2005a); Sti05: Still et al. (2005); Fyn05a: Fynbo et al. (2005a); 
Kan08: Kann et
al. (2009); Fyn06: Fynbo et al. (2005b); Wat06: Watson et al. (2006a); 
Ber05b: Berger et al. (2005b); dUP07: de Ugarte Postigo (2007);
Cen05: Cenko et al. (2005); Hol07: Holland et al. (2007);
Fol05: Foley et al. (2005);  Che05: Chen et al. (2005); Sta05: Starling et al. (2005);
DeP07: de Pasquale et al. (2007);
Fyn05d: Fynbo et al. (2005c); Sch07: Schady et al. (2007);
Pro05: Prochaska et al. (2005);
 Fyn05d: Fynbo et al. (2005f); Jak06: Jakobsson et al. (2006); 
Hil05: Hill et al. (2005); Cen06b: Cenko et al. (2006b); Mis07: Misra et al. (2007);
Fyn06: Fynbo et al. (2006a); 
Cuc06: Cucchiara Fox \& Berger (2006); Cur07: Curran et al. (2007);
Pro06: Prochaska et al.  (2006); Ell06: Ellison et al. (2006); 
Blo06: Bloom et al. (2006); Ber06: Berger \& Gladders (2006); 
Tho08: Th\"one et al. (2008) Pri06: Price Berger \& Fox (2006); Man07:
Mangano et al., (2007); 
Th\"o06: Th\"one et al., (2006); Gru07: Grupe et al. (2007); 
Fug06: Fugazza et al. (2006);  Rol06: Rol et al. (2006); 
Fyn06b: Fynbo et al. (2006b); RuV07: Ruiz-Velasco et al., (2007);
Osi06: Osip Chen \& Prochaska (2006); Mun07: Mundell et al. (2007);
Blo06: Bloom Perley \& Chen (2006), Pag07: Page et al. (2007); 
Per08a: Perley et al. (2008a); Jau07: Jaunsen et al. (2007); Tro07: Troja et al. (2007); 
Fox et al.  (2007);  Per07: Perley et al. (2007); Per08b: Perley et al. (2008b); 
Pro07: Prochaska et al. (2007); Cov08: Covino et al. (2008); 
Pro08:  Prochaska et al. (2008); PeB08: Perley \& Bloom (2008a); 
Vre08: Vreeswijk et al. (2008); Blo09: Bloom et al. (2009).
  }
\end{center}
\label{nh_av_pow}
\end{table*}  

\subsection{Broken power--law model}

In order to test for the presence of possible spectral breaks  within the
XRT energy range we selected the GRBs whose spectra have high 
signal-to-noise,  namely those which, after the applied rebinning, had a minimum of 
50 energy  bins. This choice, on average, 
corresponds to  a minimum of 1000 counts per spectrum.
We found 20 events fulfilling this  condition. 
In the excluded 13 cases (i.e. GRB 050319,  GRB 050408, GRB 050525A, GRB 050801,
GRB 050824, GRB 051111, GRB 060512, GRB 060526, GRB 060904B, GRB
060927, GRB 070125, GRB 071010A and GRB 080310) 
the spectrum in the considered time intervals has too low  S/N for fitting 
a broken power--law model  which has two more free parameters (i.e the spectral index of the 
second power-law component and the energy break between the two power-laws) 
with respect to the single power-law 
model with galactic and intrinsic absorption.

We used 2 absorption components for the broken 
power-law models, as described in the Section \ref{abp}.
The break energy $E_{\rm b}$ between the low and high energy power--laws spectral indices ($\beta_{\rm X,1}$ and $\beta_{\rm X,2}$, respectively) was left free to 
vary in the 0.3-10 keV energy range.
 Clearly a significant broken power--law fit should result in statistically different 
$\beta_{\rm X,1}$ and $\beta_{\rm X,2}$.
Therefore    no pre--determined
relation between the model parameters was assumed 
(as done for instance if the emission process is assumed to be synchrotron ($\beta_{\rm X,1}=\beta_{\rm X,2}-0.5$).

The broken power--law with a free rest frame $N_{\rm H}^{\rm host}$ 
model (hereafter ABP)  has 5 free parameters
while the absorbed single power--law model (hereafter AP) has 3 free
parameters that are a subset of the ABP model ones.   models are
nested with a progression of 2 free parameters so an ABP model fitting
is considered an improvement of the AP model one  if
$\Delta \chi^2=\chi^2_{\rm AP}-\chi^2_{\rm ABP}>4.6$ (90\% confidence).   A similar choice was also done 
by Butler \&  Kocevski (2007): they considered as acceptable a more 
complex model (with an additional free parameter) if $\Delta \chi^2 >2.7$. 
 
In 7 events (i.e. GRB 050802, GRB 050820A,  GRB 060210, GRB 060729, 
GRB 061007, GRB 061126, GRB 080319B) the fit with the ABP model resulted in an acceptable $\chi^2/{\rm dof}$ 
and the 5 free parameters of the ABP model were constrained
with acceptable uncertainties
 (i.e. a $\chi^2$ minimum is found 
inside the parameters definition range also considering their uncertainties).   
 Usually both the high energy photon index ($\beta_{\rm X,2}$) and $E_{\rm b}$ are
well constrained ( typical errors of about 0.1 for the spectral index $\beta_{\rm X,2}$ and 0.15 keV for the break energy $E_{\rm b}$ while $\beta_{\rm X,1}$) and 
$N_{\rm H}^{\rm host}$ 
 are affected by larger -- but still acceptable uncertainties (about 0.2 and 50\%, respectively)
(see Tab. 2).
 For all the 7 events the improvement of the ABP fit
with respect to the AP one  yields 
 at least a 90\% significant improvement.

In 8 cases (i.e. GRB 050318, GRB 050401, GRB 050416A, GRB 050922C,
GRB 060614, GRB 060908, GRB 070110, GRB 071003)  the ABP model  is 
not preferred to the AP one, either because 
 $\beta_{\rm X,1}$ is equal to $\beta_{\rm X,2}$ within their errors
or $E_{\rm b}$ results outside the considered energy range.   

In 5 GRBs (i.e. GRB 060124, GRB 060206, GRB 061121, GRB 050730, GRB
060418) although the $\chi^2$ of the ABP model is lower 
than that of the AP model, the improvement of the fit is not statistically significant.

 We re-analysed the spectra of these 5 events assuming an ABP model with 4 parameters,  namely  
with $N_{\rm H}^{\rm host}$ frozen to the value estimated from $A_V^{\rm host}$ 
assuming that the $A_V^{\rm host}$ -- 
$N_{\rm H}^{\rm host}$ relation reported  by Schady et al. (2007) (their Eqs. 1, 2 or 3). For
each burst we choose the conversion corresponding to the extinction curve adopted 
to obtain the  $A_V^{\rm host}$ from the analysis of its optical SED. 

In Tab. 3 we report the best X--ray spectral fit parameters values for these 5 events and the related $\chi^2_{\rm R}$. For all of these events we obtain a good fit to the data with $\chi^2_{\rm R}$ values close to unity like in the AP case.

As the AP parameters are no more a subset of the parameters of this model 
(i.e. they are not nested models) the $\Delta \chi^2$ does not provide statistical 
information on the fit improvement (e.g. Protassov et al. 2002).  
 
No a priori relation between $\beta_{X,1}$ and $\beta_{X,2}$  was assumed
and the extremely hard $\beta_{X,1}$ obtained for GRB
061121 and GRB 060418  cannot be easily accounted for by the standard 
emission processes. 
 Given the uncertainties in the inferred $N_{\rm H}^{\rm host}$, we then 
fixed or constrained the value of $\beta_{X,1}$ in two ways, assuming:
i)  the relation $\Delta\beta = \beta_{X,1}-\beta_{X,2}=0.5$;
ii) $\beta_{X,1}=0$.
In both cases the best fit returns the same $\chi^2_{\rm red}$ value.
For GRB 061121 the derived columns are $N_{\rm H}^{\rm host}=0.58^{+0.20}_{-0.13}\times 10^22$ cm$^{-2}$
($\Delta \beta =0.5$)
and $N_{\rm H}^{\rm host}=0.44^{+0.32}_{-0.13}\times 10^22$ cm$^{-2}$ ($\beta_{X,1}=0$), while 
for GRB 060418 $N_{\rm H}^{\rm host}=0.28^{+0.2}_{-0.16}\times 10^22$ cm$^{-2}$ 
($\Delta \beta =0.5$) and 
$N_{\rm H}^{\rm host}=<0.25\times 10^22$ cm$^{-2}$  ($\beta_{X,1}=0$). 
We conclude that for these two bursts 
the data cannot robustly constrain the low energy spectral slope as 
an acceptable fit can be obtained for not so extreme 
values of  $\beta_{X,1}$.
The column densities obtained  in these cases are
intermediate between the ones obtained through the ABP and the AP
models,   in agreement with what found by Schady et al. (2007) 
when assuming $\Delta \beta =0.5$.

\begin{table*}
\begin{center}
\begin{tabular}{llllllll}
\hline
\hline
GRB		&$z$&$N_{\rm H}^{\rm host}$&$\beta_{\rm X,1}$& $E_{\rm b}$ & $\beta_{\rm X,2}$&  $\chi^2_{\rm R}$ (dof)& prob\\ 
	  	  &     & $10^{21}$ cm$^{-2}$   &                             & keV               &                            &    \\
\hline
\hline
050802  &0.55   &0.6$\pm$0.6         &0.58$^{+0.13}_{-0.14}$ &1.64$^{+0.63}_{-0.64}$ &0.95$\pm$0.12 &0.99 (157)& $6.5e^{-3}$\\
050820A &2.612  &2.2$^{+2.2}_{-2.2}$ &0.63$^{+0.15}_{-0.20}$ &1.05$^{+0.70}_{-0.33}$ &1.00$\pm$0.07 &0.947 (141)&$7.7e^{-2}$ \\
060210  &3.91   &4.$^{+7}_{-4}$      &0.59$^{+0.32}_{-0.22}$ &1.15$^{+0.23}_{-0.17}$ &1.12$\pm$0.07 &0.99 (183)& $8.5e^{-2}$ \\ 
060729  &0.54   &0.$^{+0.2}_{-0}$    &0.53$^{+0.24}_{-0.09}$ &1.13$^{+0.13}_{-0.10}$ &1.04$\pm$0.04 &0.97 (288)& $2.9e^{-3}$\\
061007  &1.26   &3.0$^{+0.9}_{-0.9}$ &0.02$^{+0.36}_{-0.34}$ &0.80$^{+0.04}_{-0.05}$ &0.86$\pm$0.02 &1.02 (478)& $3.9e^{-4}$\\
061126  &1.1588 &1.7$^{+3.6}_{-1.7}$ &-0.16$^{+0.82}_{-1.2}$ &1.05$^{+0.27}_{-0.21}$ &0.74$\pm$0.08 &1.056 (141)&$9.8e^{-2}$ \\
080319B &0.937  &0.7$^{+0.2}_{-0.2}$ &0.49$^{+0.08}_{-0.10}$ &1.14$^{+0.08}_{-0.08}$ &0.81$\pm$0.01 &1.27 (608)& $8.6e^{-9}$ \\
\hline
\hline
\end{tabular}
\caption{Results of the fit to the X--ray spectra with the absorbed broken power--law for the 7
   bursts for which the model parameters are constrained. 
Note that $\beta$ represent the energy spectral index ($\beta=\Gamma-1$).
The analysed spectra have been extracted in time intervals as
in the third column of Tab. \ref{nh_av_pow}. 
}
\end{center}
\label{broken_libera}
\end{table*}  

\begin{table*}
\begin{center}
\begin{tabular}{lllllll}
\hline
\hline
GRB	     &$z$&$N_{\rm H}^{\rm host}$&$\beta_{\rm X,1}$& $E_{\rm b}$ & $\beta_{\rm X,2}$&  $\chi^2_{\rm R}$ (dof)\\ 
	     &     & $10^{21}$ cm$^{-2}$   &                          & keV               &                            &     \\
\hline
\hline
050730 &3.967&0      &0.36$^{+0.23}_{-0.27}$ &1.00$^{+0.6}_{-0.25}$   & 0.75$^{+0.09}_{-0.09}$& 1.22 (94)\\
060206 &4.045&0      &0.03$^{+0.62}_{-1.4}$   &0.63$^{+0.18}_{-0.11}$ & 1.23$^{+0.12}_{-0.12}$&0.98 (86)\\
060124 &2.296& 0     &0.47$^{+0.14}_{-0.19}$ &1.27$^{+0.27}_{-0.25}$ & 1.05$^{+0.1}_{-0.1}$    & 0.81 (106)\\
060418 &1.489& 0     &-0.23$^{+0.5}_{-0.6}$   &0.79$^{+0.2}_{-0.09}$    & 0.81$^{+0.08}_{-0.07}$&0.85 (90)\\ 
061121 &1.314&1.44 &-0.89$^{+0.46}_{-0.65}$&0.79$^{+0.09}_{-0.08}$ & 0.89$^{+0.06}_{-0.06}$&0.88 (120)\\
\hline
\hline
\end{tabular}
\caption{
Results of the absorbed broken power--law model fitting obtained by freezing 
   the value of $N_{\rm H}^{\rm host}$  to that estimated from $A_V^{\rm host}$ 
   through Eq. 1, 2 or 3 in  Schady et al. (2007) (see text). 
The analysed spectra have been extracted in time intervals as 
in the third column of Tab. \ref{nh_av_pow}.  
}
\end{center}
\label{broken_fixed}
\end{table*}   

\subsection{Discussion on the X--ray spectral analysis}
\label{nhav}
The breaks that we have found are all in the range between 0.6 and 1.6 keV. This  is likely due to the fact that 
the peak of the  effective area of the Swift-XRT is 1.5 keV and it is therefore easier to find a spectral break when it falls around this energy. A break at $E_{\rm b}<0.6$ keV is hardly detectable and therefore we cannot exclude its presence in most of spectra. Also a break at  $E_{\rm b} >3$ keV can not be easily detectable with the available late time X--ray spectra but we expect in this case to obtain a quite hard spectrum fitting with an AP model (i.e. a single power law spectral index with a value similar to the obtained  $\beta_{\rm X,1}$). Since the values of  $\beta_{\rm X}$ reported in Tab.  \ref{nh_av_pow} are usually not so hard we do not expect we are missing a large number of  $E_{\rm b} >3$ keV breaks.



\begin{figure}
\vskip -0.7cm
\hskip -0.7 cm
\psfig{figure=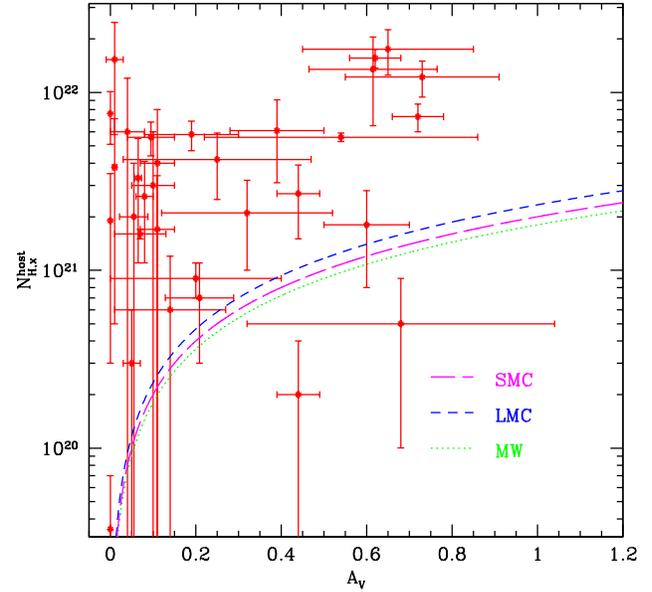,width=9.7cm,height=9.7cm}
\vskip -0.8cm
\caption{
Rest frame column densities $N_{\rm H}^{\rm host}$  
(obtained from fitting a single power--law model to the X--ray data)
versus the visual extinction $A_V^{\rm host}$ in the GRB host galaxy for all 33
GRBs of the sample. 
The three curved lines correspond to
the $N_{\rm H, x}$ versus $A_V$ relations observed in the Milky Way 
and in the Small Magellanic Cloud as described by Eqs. 1, 2 or 3 in Schady et al. (2007).
} 
\label{nh_av_powerlaw}
\end{figure}

It has been already pointed out (e.g. Galama \& Wijers 2001, Stratta et
al. 2004, Schady et al. 2007,  Starling et al. 2007, Watson et al. 2007) that the  $N_{\rm H}^{\rm host}$ derived  
from the fit of the X--ray spectra
are usually quite large (up to a few $\times 10^{22}$
cm$^{-2}$). On the other hand the values inferred for host galaxy dust reddening $A_V^{\rm host}$ are usually small. This is true also for the GRBs in our sample as shown in  Fig. \ref{nh_av_powerlaw}. 



\begin{figure}
\vskip -0.7cm
\hskip -0.7 cm
\psfig{figure=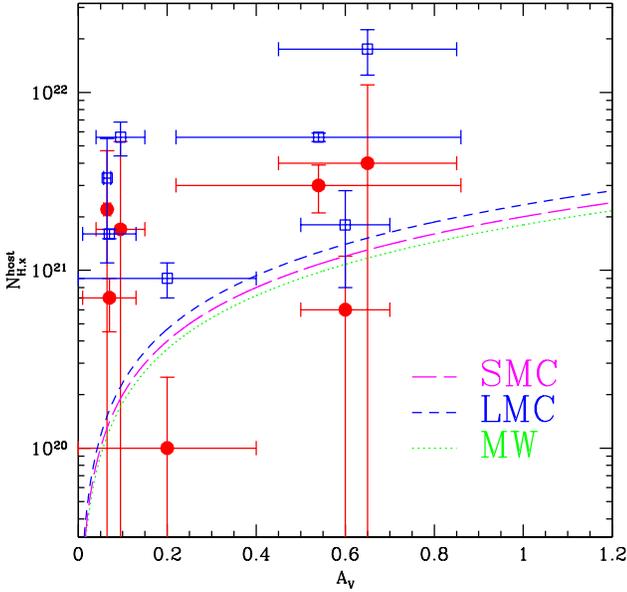,width=9.7cm,height=9.7cm}
\vskip -0.8cm
\caption{
Rest frame column densities $N_{\rm H}^{\rm host}$ versus 
visual extinction $A_V^{\rm host}$ for the 7
GRBs in which a broken power--law model gave an acceptable fit (see text).
Filled circles represent the column densities obtained from
an absorbed broken power law fit to the XRT spectra with the local absorption fixed  to the Galactic 
$N_{\rm H}^{\rm Gal}$ values while empty squares  indicate
the $N_{\rm H}^{\rm host}$ obtained from a single power--law fitting 
for the same events.
The three curved lines represent the $N_{\rm H, x}$ versus
$A_V$ relations observed in the Milky Way and in the Small
Magellanic Cloud as described 
in Schady et al. (2007).
} 
\label{nh_av_libera}
\end{figure}

From the observational point of view  the  large $N_{\rm H}^{\rm host}$ 
derived from the fitting of the X--ray spectrum corresponds to a 
deficit of counts below approximately 1 keV with respect to the 
extrapolation of a single power--law model.
In principle, this deficit could instead be due to an intrinsically curved
or a broken power--law spectrum. 
For the 7 GRBs for which the ABP model gave a better fit (with respect to the AP model)
we can check if the obtained values of $N_{\rm H}^{\rm host}$ are in agreement
with the optical extinction assuming a $N_{\rm H, X}$ vs. $A_V$ relation  (e.g. Eqs. 1, 2 or 3 in Schady et al. 2007). $A_V^{\rm host}$. 
Fig. \ref{nh_av_libera} shows the values of $N_{\rm H}^{\rm host}$
obtained  with  the ABP model fitting  versus $A_V^{\rm host}$ (filled
circles).
 For comparison, also the 
$N_{\rm H}^{\rm host}$ values obtained with the AP model fitting  (empty squares) are reported. 
The solid lines represent the Milky Way and Small Magellanic Cloud like relations as in Fig. \ref{nh_av_powerlaw}.
For 5 GRBs the uncertainties on $N_{\rm H}^{\rm host}$ are quite large, 
making these values consistent with zero,  i.e. they must be considered as upper limits.
These limits, always smaller than the $N_{\rm H}^{\rm host}$ values
obtained with the AP model, are consistent with the
observed $A_V^{\rm host}$. 
For the remaining 2 GBRs (GRB 060210 and 080319B)  however the
value of $N_{\rm H}^{\rm host}$ are still somewhat larger than what
expected by the standard gas-to-dust relation, though clearly the disagreement 
with is less pronounced.


 While the presence of an intrinsic break in the emitted X--ray
spectrum can solve or mitigate the problem of an excess of $N_{\rm H}^{\rm host}$ 
with respect to the optical reddening for a fraction
of events,  this can not be considered as a general solution of this issue, 
on the basis of different indications.

 As the excess is
observed in a large fraction of GRBs, 
this would imply that the observed X--ray spectrum is almost always
a broken power law, with a break in the rather narrow 0.5--1.5 keV
energy range, even if the redshifts of these bursts are different.

\begin{figure}
\vskip -0.7cm
\hskip -0.7cm
\psfig{figure=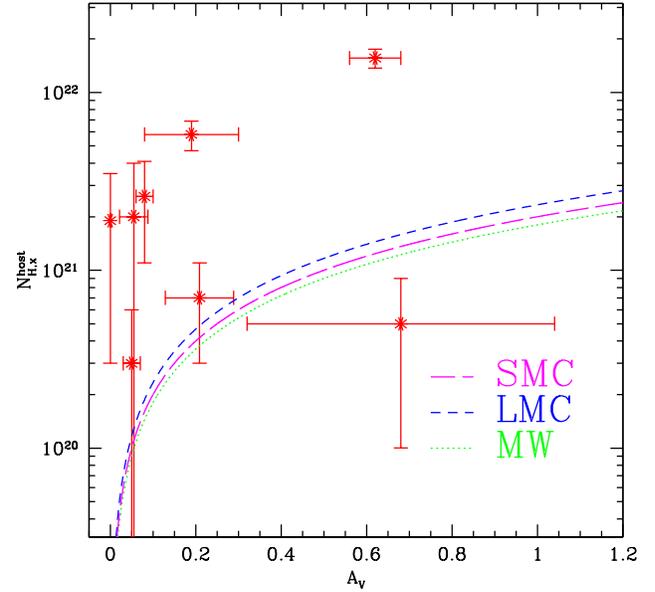,width=9.7cm,height=9.7cm}
\vskip -0.8cm
\caption{
Rest frame column densities $N_{\rm H}^{\rm host}$ versus the
visual extinction $A_V^{\rm host}$ obtained from the AP model for the 
events where the ABP  model is excluded.
The two curved lines represent the $N_{\rm H, x}$ versus
$A_V$ relations observed in the Milky Way and in the Small
Magellanic Cloud as described 
in Schady et al. (2007).} 
\label{nh_av_pow_nobreak}
\end{figure}

Furthermore  we can directly exclude the presence of a
spectral break inside the observed XRT spectrum for about half of the
analysed events. 
In general, these events have an intermediate/high $N_{\rm H}^{\rm host}$ (when fitted with the AP model; see
Fig. \ref{nh_av_pow_nobreak} compared to Fig. \ref{nh_av_libera}). 


\section{Optical to X--rays SEDs}
\label{sed}

In this section we present the broad band SEDs for the 7  events in which we found a break in the X--ray spectrum  to check whether 
the spectra, when extrapolated to lower frequencies, are consistent with 
the available optical photometry at the same epoch. Then the 
optical to X--ray SED information will be combined with the results on the decomposition 
of the light curves behaviour in two components, as suggested by G09.

\begin{figure}
\vskip -0.9cm
\hskip -1.2cm
\psfig{figure=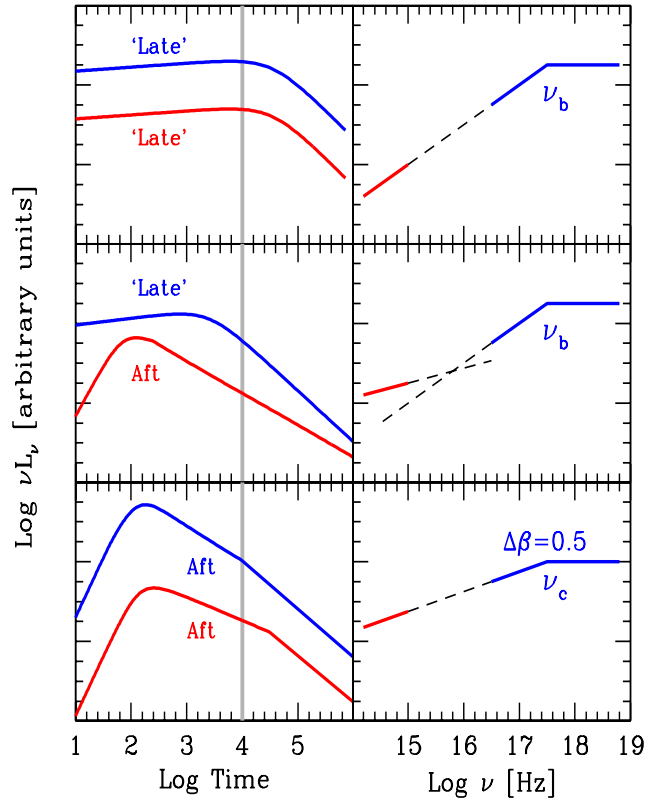,width=11cm,height=13cm}
\vskip -1.5 cm
\caption{Sketch illustrating the possible different  cases for the relation between 
light curves behaviour in terms of the two component decomposition and SEDs.
The left panels refer to the X--ray (upper curves) and optical 
(lower curves) light curves, while  the corresponding expected 
optical to X--ray SED are shown in the right-hand panels.
The bottom right panel shows the standard ``afterglow--afterglow" case, i.e. both 
light curves are dominated by the afterglow component, with a cooling break 
appearing first at X--ray energies. 
The vertical grey line  indicates the time of the extraction of the SED.
$\nu_{\rm b}$ and $\nu_{\rm c}$ represent the break and cooling frequency, respectively. 
}
\label{combo6}
\end{figure}

A key test of the  two component light curve modelling by G09 is to verify whether the spectral properties at different
times are consistent with what inferred by the light curve de-convolution. The 7 GRBs 
whose X-ray spectrum is  consistent with the presence of a break in the X--ray band are the best candidates to perform  the test  as shown in fig. \ref{combo6}.

For each burst we select epochs where simultaneous optical photometry and XRT 
observations are available in order to  use the most complete spectral information available
and to avoid (if possible) flux extrapolations.
This limits the number of optical to X--ray SED considered.

The X--ray spectrum  is extracted from a time interval around 
the selected epoch in order to have at least 50 energy bins and it is 
re--normalised to the 0.3--10 keV flux obtained from  the light curve.  
The spectral index plotted in the SEDs are the ones reported in Tab. 2. 
As done in G09 we used the light curves from the {\it Swift} repository (see Evans et al. 2009). The counts to flux conversion instead was done using the values from our broken power law spectral fits.
When the optical and X--ray bands light curves track each other, 
 i.e. they are dominated by the same component, one 
single SED is considered; when instead  they show different temporal 
behaviours we considered more SEDs to test the modelling at different phases of the evolution of 
the two components. 

In the following we  present the results for 6 of the 7 GRBs separately.
The complexity of the remaining one (i.e. GRB 080319B) prompted us to discuss it in the details in a dedicated paper (Nardini et al. 2009 in preparation).
We anticipate that no event shows an optical to X--rays SED that
is inconsistent with the presence of a break in the XRT band and the two 
component interpretation.

\subsection{GRB 050802}

\begin{figure}
\vskip -0.5cm
\psfig{figure=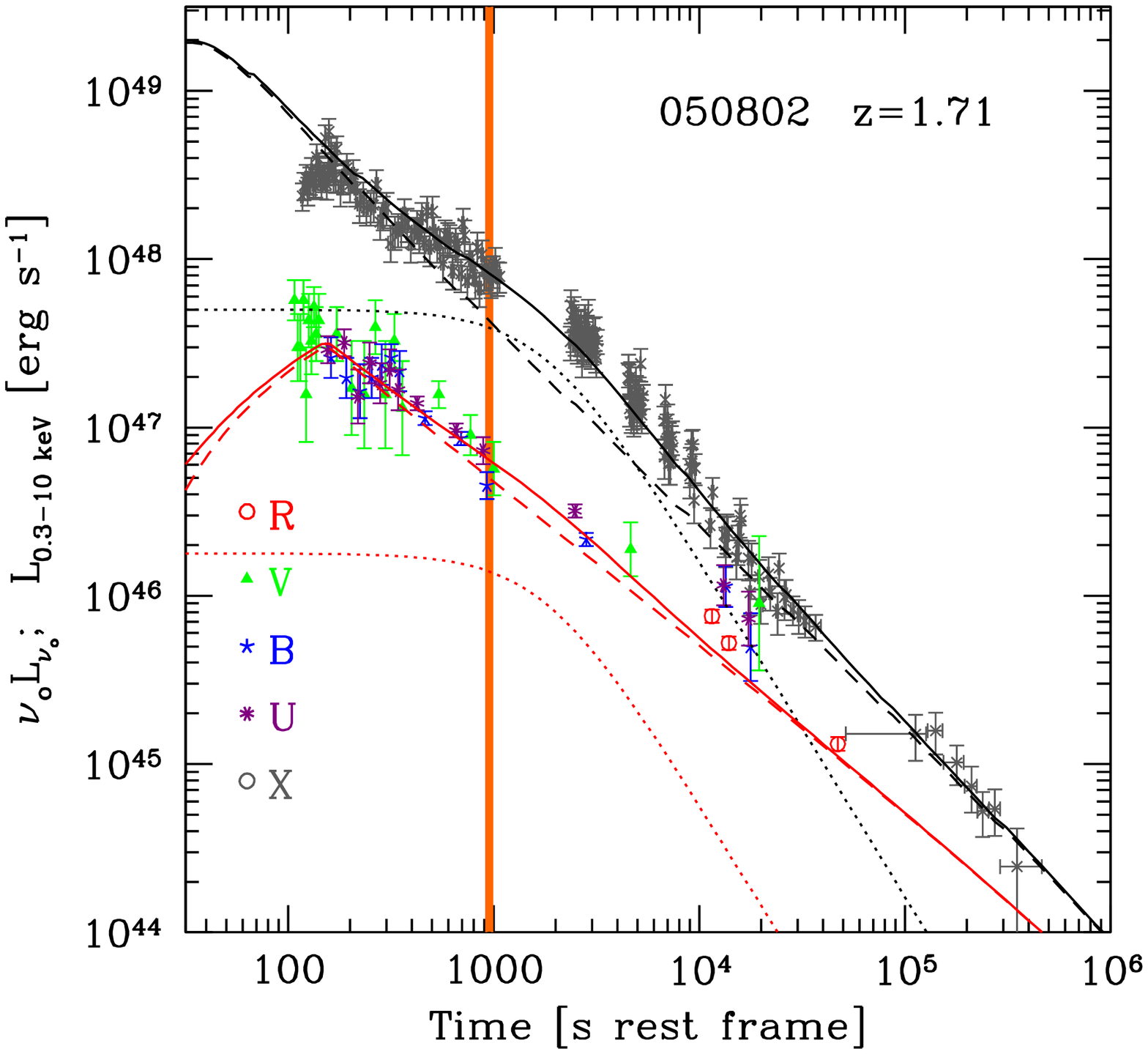,width=9cm,height=7.5cm} 
\vskip -0.5 cm
\psfig{figure=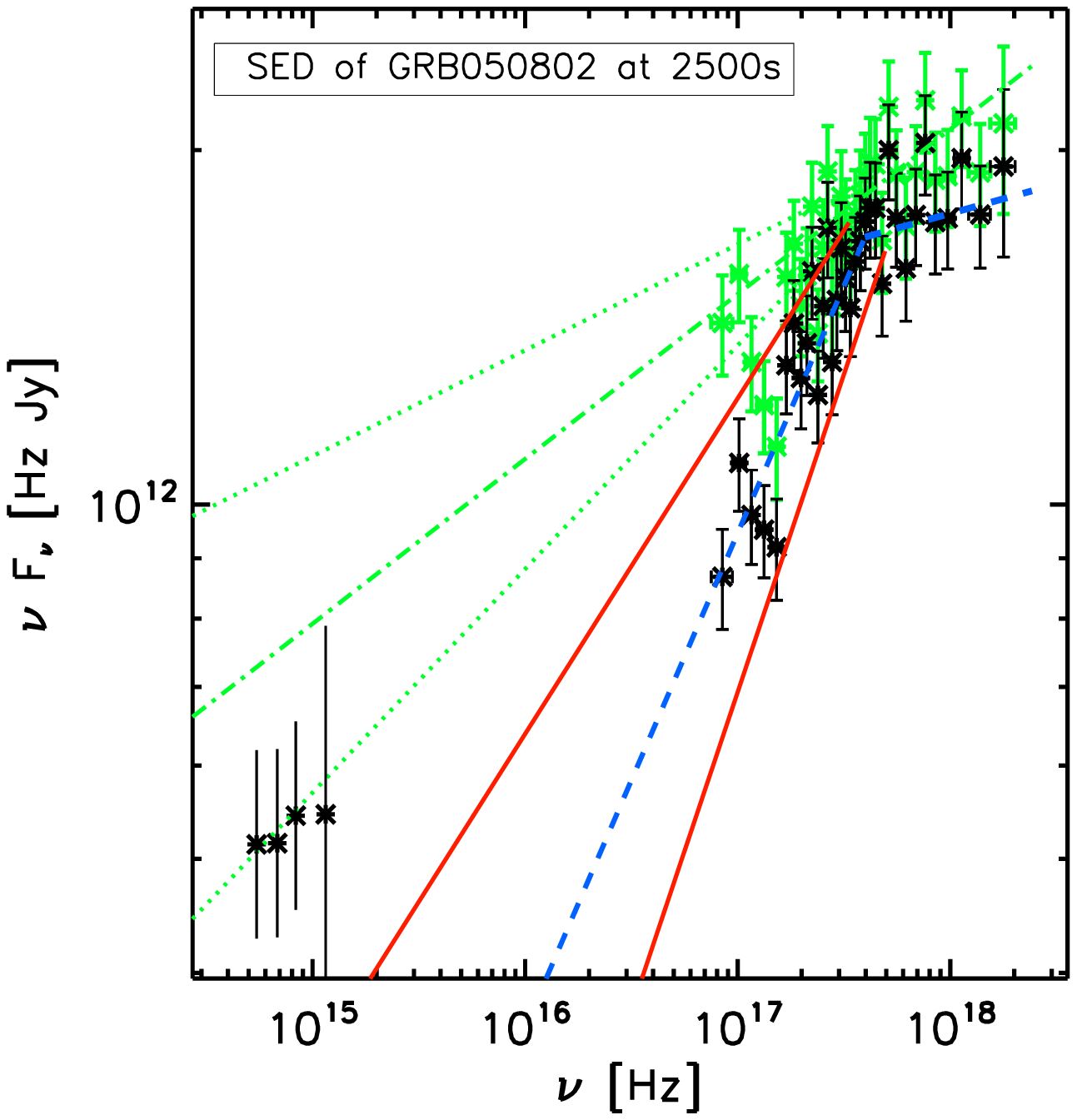,width=8.5cm,height=7.cm}
\caption{
Top panel:
X--ray (in grey) and optical (different symbols, as labelled)
light curves of GRB 050802  (in the rest frame time). 
Lines indicate the model fitting: afterglow  component (dashed line),
``late prompt" one (dotted line) and their sum (solid line). 
Black lines refer to the X--rays, light grey (red in the electronic 
version)  to  the optical. 
 The vertical line marks the time at which the SED is extracted.
Bottom panel:
Optical to X--ray $\nu F_{\nu}$ SED  at about  2500 s 
(observed time, corresponding to 920 s rest frame)
after trigger. }
The  dashed and solid lines show respectively the best fit (with the ABP model) to the X--ray spectrum (the spectral parameters are reported in Tab. 2) and the uncertainties  on the slope of the low energy spectral index 
$\beta_{\rm X, 1}$. The dotted line shows  the best fit (with the AP model) to the X--ray spectrum.

\label{050802lc}
\end{figure}

 The optical light curve photometric data are mainly from Oates et al. (2007) together with 
later time $R$ band data from GCNs (Pavlenko et al. 2005, Fynbo et al. 2005). 
The {\it  Swift} UVOT filters  uvm2 and uvw2 are strongly affected by Ly$\alpha$ dumping 
and are not considered. 
The X--ray light curve  has been modelled  as the combination of ``standard afterglow" 
emission dominating at early and at late times (before 700 s and after about  
10 ks, rest frame) and ``late prompt" emission dominating   in between. 
 The ``standard afterglow"  component instead describes the evolution of the 
optical flux during the whole period of  the follow up (see Fig. \ref{050802lc}).
 
We extracted the optical to X--ray SED around an observed time of  2500s (920 s rest frame) 
when the optical light curve is dominated by the standard afterglow and ``late prompt" emission is becoming predominant in the X--rays.
 The X--ray spectrum represented in Fig. \ref{050802lc} has been extracted in the time interval reported in Tab. 1.
Schady et al. (2007)  estimated  a non negligible host galaxy dust absorption 
($A_{\rm V}^{\rm host}=0.55\pm 0.1$) on the basis of a Milky Way extinction curve 
and assuming a power--law spectrum connecting the optical and X--ray bands. 
By  considering  the optical bands alone we find a similar  $A_{\rm V}^{\rm host}=0.6$ 
with an optical spectral index $\beta_{\rm o}\approx 0.9$.  

The SED, plotted in Fig. \ref{050802lc}, is consistent with the optical 
and the X--ray emission being 
dominated by different components  (note that  also Oates et al. 2008 and de Pasquale et al. 2009 found a similar inconsistency) with a spectral break falling in the observed 
XRT energy range, as  indeed obtained from the X--ray spectral analysis. 
Note that the 
X--ray spectra shown in this section  have  been ``de-absorbed" both for the 
galactic and the host frame (when present) contributions.

\subsection{GRB 050820A}

\begin{figure}
\vskip -0.5cm
\psfig{figure=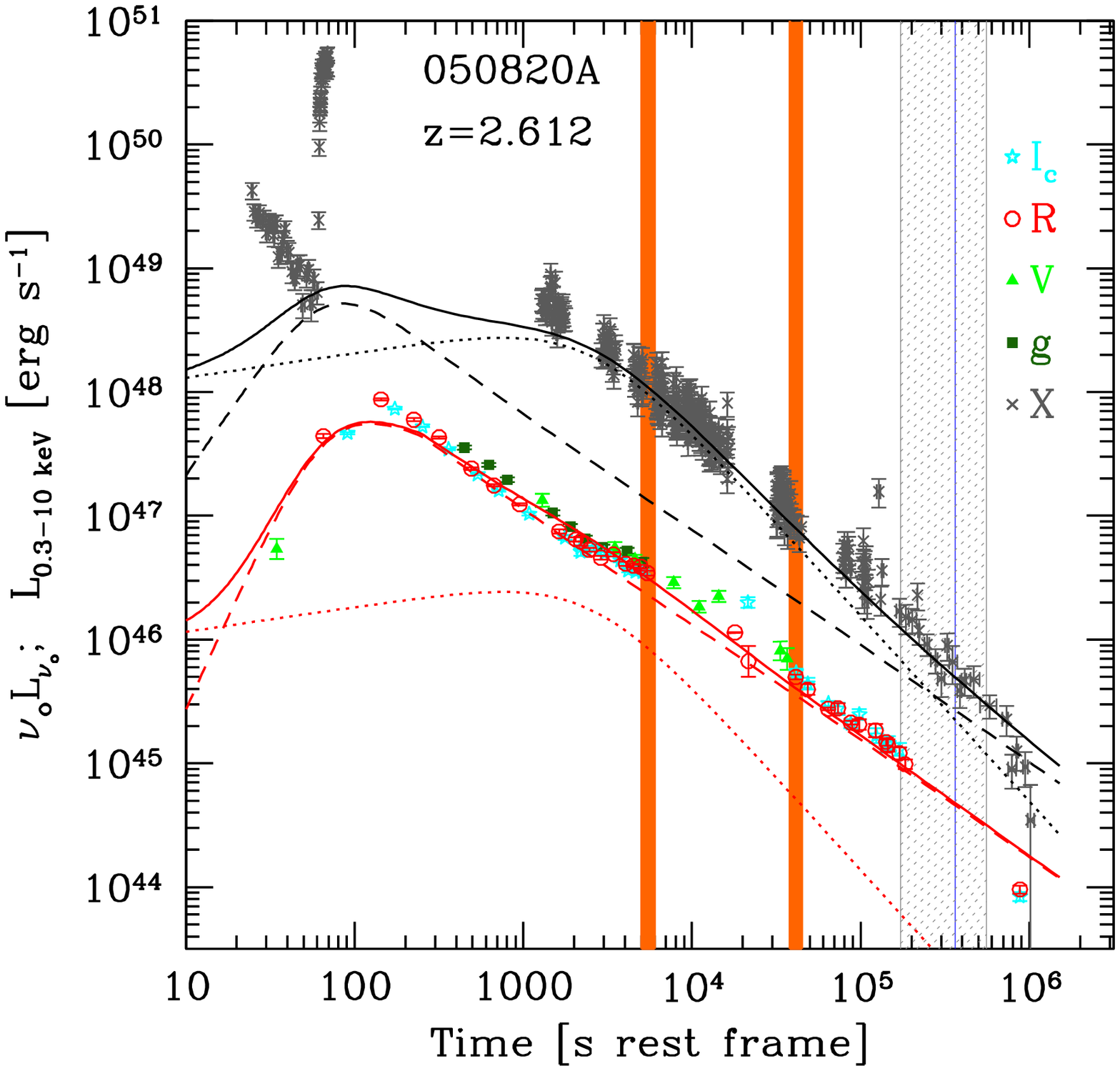,width=9.cm,height=7.5cm} 
\vskip -0.5 cm
\psfig{figure=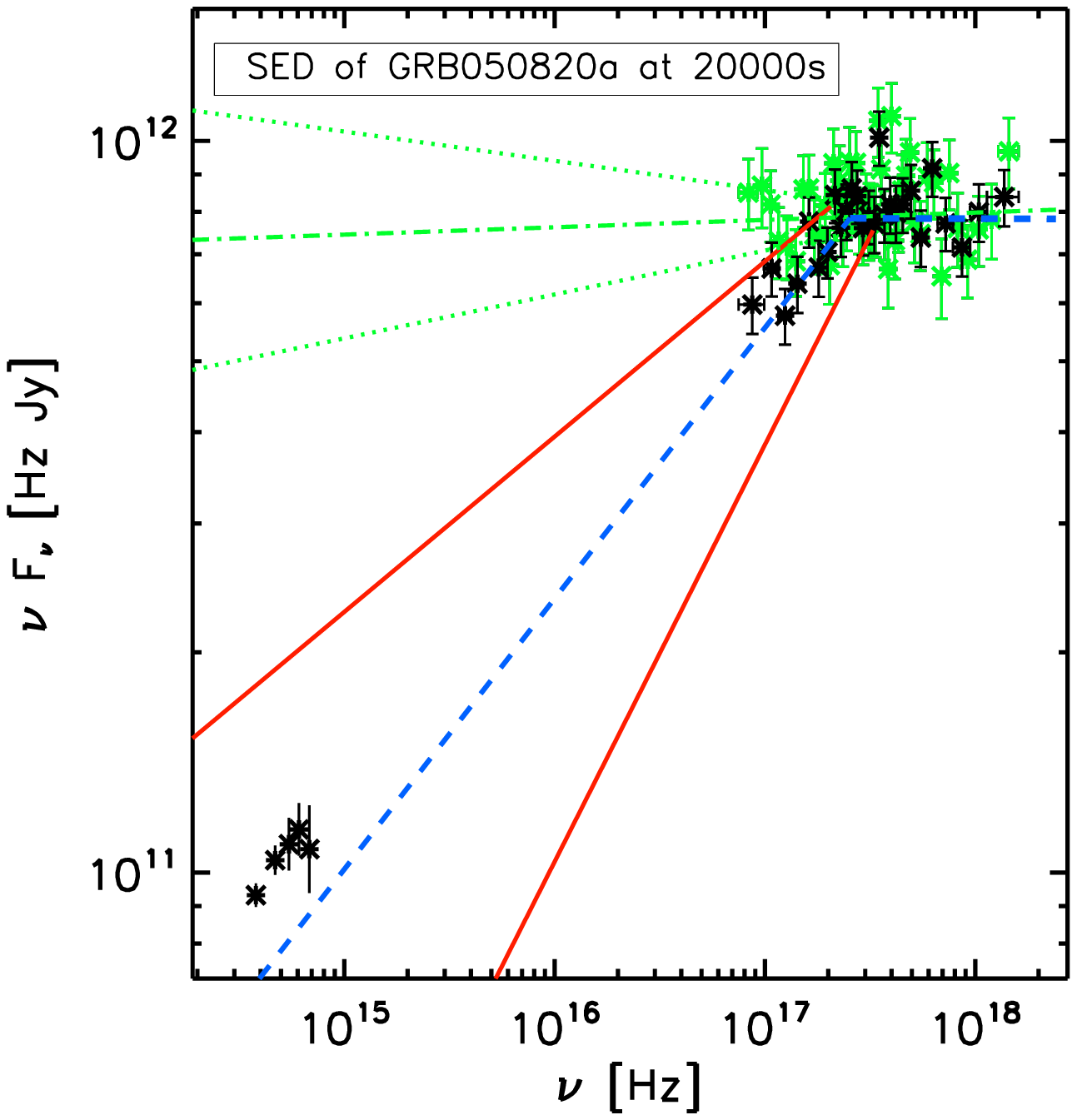,width=8.5cm,height=7cm}
\vskip -0.1 cm
\psfig{figure=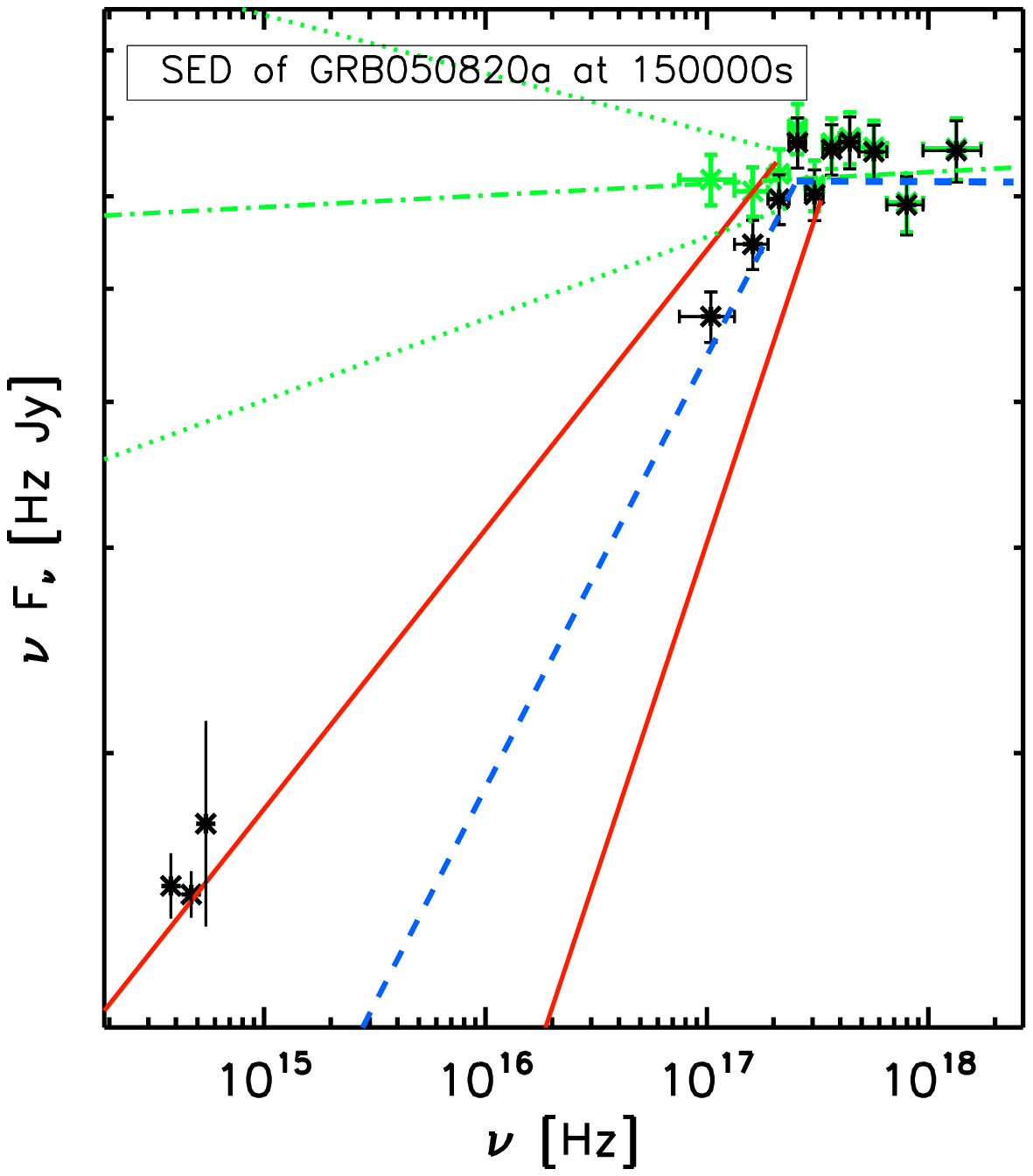,width=8.5cm,height=7cm}
\caption{
Top panel: X--ray 
and optical 
light curves of GRB 050820A (rest frame time). 
Same notation as in Fig. \ref{050802lc}. 
The thin vertical line represents the jet break time with  its estimated errors (see Ghirlanda et al. 
2007 and references therein).  The thick vertical lines mark the times at which the SEDs are extracted.
Middle and bottom panels:
Optical to X--ray $\nu F_{\nu}$ SED  around 20 ks (middle) and  150 ks (bottom)
after trigger in the observer frame (corresponding to 5.5 ks and 41.5 ks
in the rest frame). 
The dashed and solid lines show respectively the best fit (with the ABP model) to the X--ray spectrum (the spectral parameters are reported in Tab. 2) and the uncertainties on the slope of the low energy spectral index 
$\beta_{\rm X, 1}$. The dotted line shows  the best fit (with the AP model) to the X--ray spectrum.
} 
\label{050820alc}
\end{figure}

 The photometric data are from Cenko et al. (2006a, 2009). 
{\it Swift}/BAT triggered on a precursor about 200 s before the main event 
(Cenko et al 2006a, Burlon et al. 2009). 
Our reference time is set at the trigger time and we do not consider the prompt X--ray emission 
 detected before the end of the main $\gamma$--ray event. 
The X--ray light curve is dominated by the  ``late prompt" component up to 200 ks 
(720 ks in the observer frame),  and by a ``standard afterglow" component after then.
The ``standard afterglow"  emission instead prevails during the entire duration of 
the optical light curve but in a time interval around 5 ks (18 ks observer frame) 
where its contribution becomes comparable with the ``late prompt" one (see Fig. \ref{050820alc}). 

We extracted two SEDs in order to  test the modelling at two different light curve phases.
The first one at about 20 ks in the observer frame ($\sim$ 5500 s rest frame) where the ``late prompt" 
gives the maximum contribution in the optical light 
curve and the available photometry is richer ($I_c$, $R_c$, $V$, $g$ and $B$ bands).  
Cenko et al. (2006a)  estimated a $\beta_{\rm o}= 0.77$ with negligible host galaxy dust 
absorption while Kann et al. (2009)  inferred an   $A_{\rm V}^{\rm host}=0.065\pm 0.008$. 
We used the latter estimate and obtained $\beta_{\rm o}\approx 0.7$. 
This first SED is plotted in Fig. \ref{050820alc} and 
shows that the optical flux lies slightly above the extrapolation of the broken power law that
 best describes the XRT spectrum,  but as the uncertainties on $\beta_{\rm X, 1}$ are quite large
the optical flux is fully consistent with the extrapolation.  In this SED the X--ray data are extracted from the time interval reported in Tab.1.

As mentioned at 5500 s (rest frame) the optical flux is due to a similar 
contribution of the ``standard afterglow" and the ``late prompt" component. The 
the cooling frequency is already redward of the considered optical bands and the 
``standard afterglow" has $\beta_{\rm o}=0.92$ (corresponding to an emitting particle 
distribution with slope $p=1.85$).  In the  ``late prompt" component modelling, the 
low energy spectral index is instead $\beta_{\rm o}=0.45$ (see Eq. 3 in G09),  
consistent within errors with $\beta_{\rm X, 1}$.
 The intermediate value of the observed optical slope is thus consistent with the predictions 
of the two component modelling. 

We  considered a second SED at about 150 ks after the trigger
(observer frame, corresponding to 41 ks rest frame).  The plotted X--ray data are from the time integrated spectrum of the complete second XRT observation. 
In this phase the X--ray light curve is dominated by the ``late prompt" while the 
``standard afterglow" dominates the optical emission. 
The combined SED is plotted in the bottom panel of Fig. \ref{050820alc} and confirms 
the proposed scenario:  the optical data are at this time much brighter than what 
predicted by  the extrapolation (with  slope $\beta_{\rm X, 1}$) of
the X--ray spectrum  to the optical bands . 
Even  though  at these late times there are only 3 available photometric points 
($I_c$, $R_c$ and $V$ band) and  the $V$ band flux is 
affected by a large error, the optical SED 
is well fitted by a softer $\beta_{\rm o}=0.95$ 
that is closer to the value predicted for the ``standard afterglow" component.

\subsection{GRB 060210}

\begin{figure}
\vskip -0.5cm
\psfig{figure=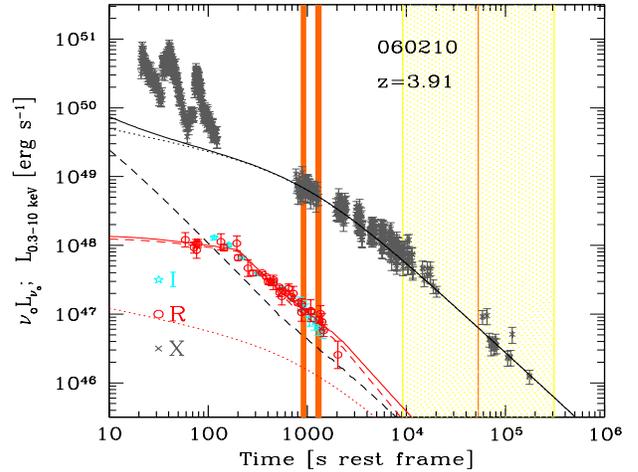,width=9cm,height=7.5cm} 
\vskip -0.5cm
\psfig{figure=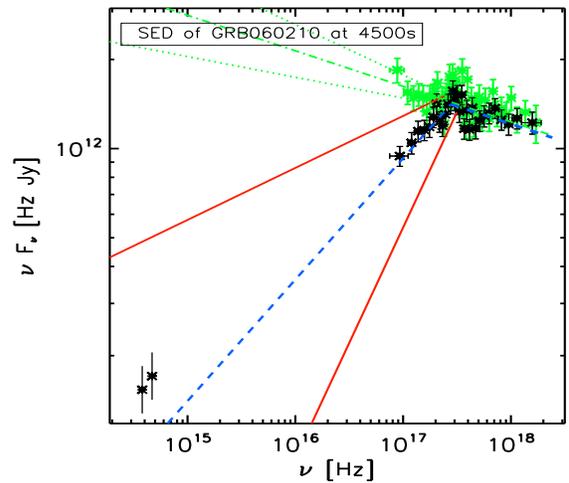,width=8.5cm,height=7cm}
\vskip -0.1cm
\psfig{figure=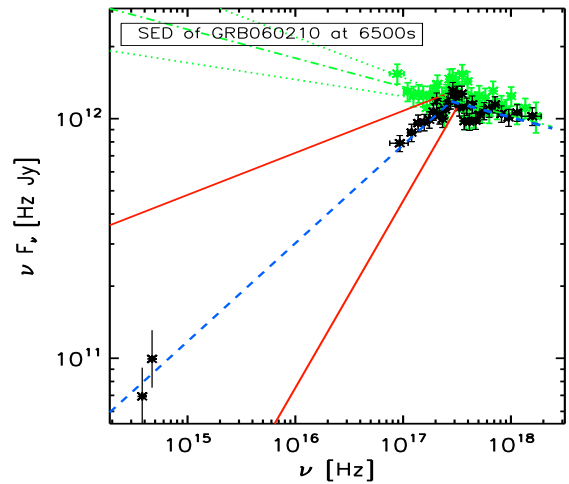,width=8.5cm,height=7cm} 
\caption{
Top panel:
X--ray (in grey) and optical (different symbols, as labelled)
light curves of GRB 060210 (in the rest frame time).
Same notation as in Fig. \ref{050802lc}.
The vertical line and stripes indicate the jet times expected if the burst followed the 
$E_{\rm peak}$ vs. $E_{\gamma}$ ``Ghirlanda relation" (Ghirlanda et al. 2007) (see text).
 The vertical lines mark the time at which the SEDs are extracted.
Mid (Bottom)  panel:
Optical to X--ray $\nu F_{\nu}$ SED of GRB 060210 
at about  4500 (6500) s after trigger in the observer frame (916 (1300) s rest frame). 
The  dashed and solid lines show respectively the best fit (with the ABP model) to the X--ray spectrum (the spectral parameters are reported in Tab. 2) and the uncertainties on the slope of the low energy spectral index 
$\beta_{\rm X, 1}$. The dotted line shows  the best fit (with the AP model) to the X--ray spectrum.
} 
\label{060210lc}
\end{figure}

 The optical afterglow has been observed in the $R$ and $I$ bands (photometric data from Curran et al. 2007), 
while, because of the high redshift, smaller wavelengths bands are not observable due to the 
Ly$\alpha$ limit. 
The X--ray light curve shows an intense flaring activity at early times 
and it is dominated by the ``late prompt"  component at later times.  
The optical light curve is sampled 
only up to $\sim$2000 s rest frame ($\sim$ 9800 s observer frame) and is 
dominated by the ``standard afterglow" emission, as shown in Fig. \ref{060210lc}. 
Since  fluxes in only two optical bands are available it is not possible to  
infer the value of $A_{\rm V}^{\rm host}$ from the optical photometry. 
Curran et al. (2007) found a very soft observed spectrum,
 after correction for Galactic extinction and 
Ly$\alpha$ absorption: the optical spectral index $\beta_{\rm o}^{\rm obs}=3.1\pm 0.4$ 
at 5000 s (observer frame).
Assuming that the optical and X--ray emission are 
produced by the same mechanism they inferred
two possible values of 
the host galaxy dust extinction, 
assuming either a single or a broken power--law joining the optical and the X--ray data.

We extracted 2 SEDs at 4500 and 6500 s 
(observer frame; see Fig. \ref{060210lc}).  In both SEDs the X--ray data are extracted from the time interval reported in Tab. 1.
In  the two component modelling  the optical and X--ray emission would be due to different 
components, to account for their different temporal behaviours. 
Thus  $A_{\rm V}^{\rm host}$ cannot be inferred from the optical to X--rays SED.  This argument, together with the fact that only 2 optical bands have been sampled, does not allow a direct dust absorption estimate, therefore we do not have constraints also on the optical spectral index.  If the optical emission is produced by a standard afterglow mechanism, we can choose as an example a value of $\beta_{\rm o}\approx 0.5$. 
If we assumed a $\beta_{\rm o}\approx 0.5$ then $A_{\rm V}^{\rm host}\approx 0.65$.  Since only two optical bands are available, the uncertainties  are very large but since this value is  
similar to the mean Small Magellanic Cloud--like $E_{\rm B-V}=0.27$ obtained by Curran et al. (2007)
in the broken power--law case the latter value is used in the correction applied to the plotted SEDs. The large error on  $\beta_{\rm X, 1}$, and
the  paucity of photometric data do not allow to draw any firm conclusion on 
this burst. 
\\
 The large uncertainties on the optical to X--ray SED at 4500 and 6500 s (observer frame), and the choice of the same  $E_{\rm B-V}=0.27$ used by Curran et al. (2007) make them  consistent with an unique broken--power law but the presence of two components cannot be excluded.


No break is observed in the X--ray light curve.  In particular, one can estimate the 
expected jet break time if the GRB was to follow the so called ``Ghirlanda relation"  (Ghirlanda et al. 2007).
However the lack of evidence for such a break is consistent with the light--curve modelling
as the X--ray flux is indeed dominated by the ``late prompt'' emission at the 
time when the jet break is expected. 
No data are available at such a time in the optical band 
where  such a break  should have been detectable, due to the dominance of the ``standard afterglow" component.

 In conclusion, the poor optical photometry of GRB 060210 does not allow to obtain an estimate of both $A_{\rm V}^{\rm host}$ and $\beta_{\rm o}$. This fact, together with the quite large errors in the $\beta_{\rm X, 1}$ estimate, does not allow to infer firm conclusions on this GRB. The diversity of the optical and X--ray temporal behaviour and the lack of jet break in the late time X--ray observations prompt us to model the light curves as due to different components.  The optical to X--ray SEDs cannot give better constraints to the model since within errors are consistent with both having one or two separate components.

\subsection{GRB 060729}

\begin{figure}
\vskip -0.5cm
\psfig{figure=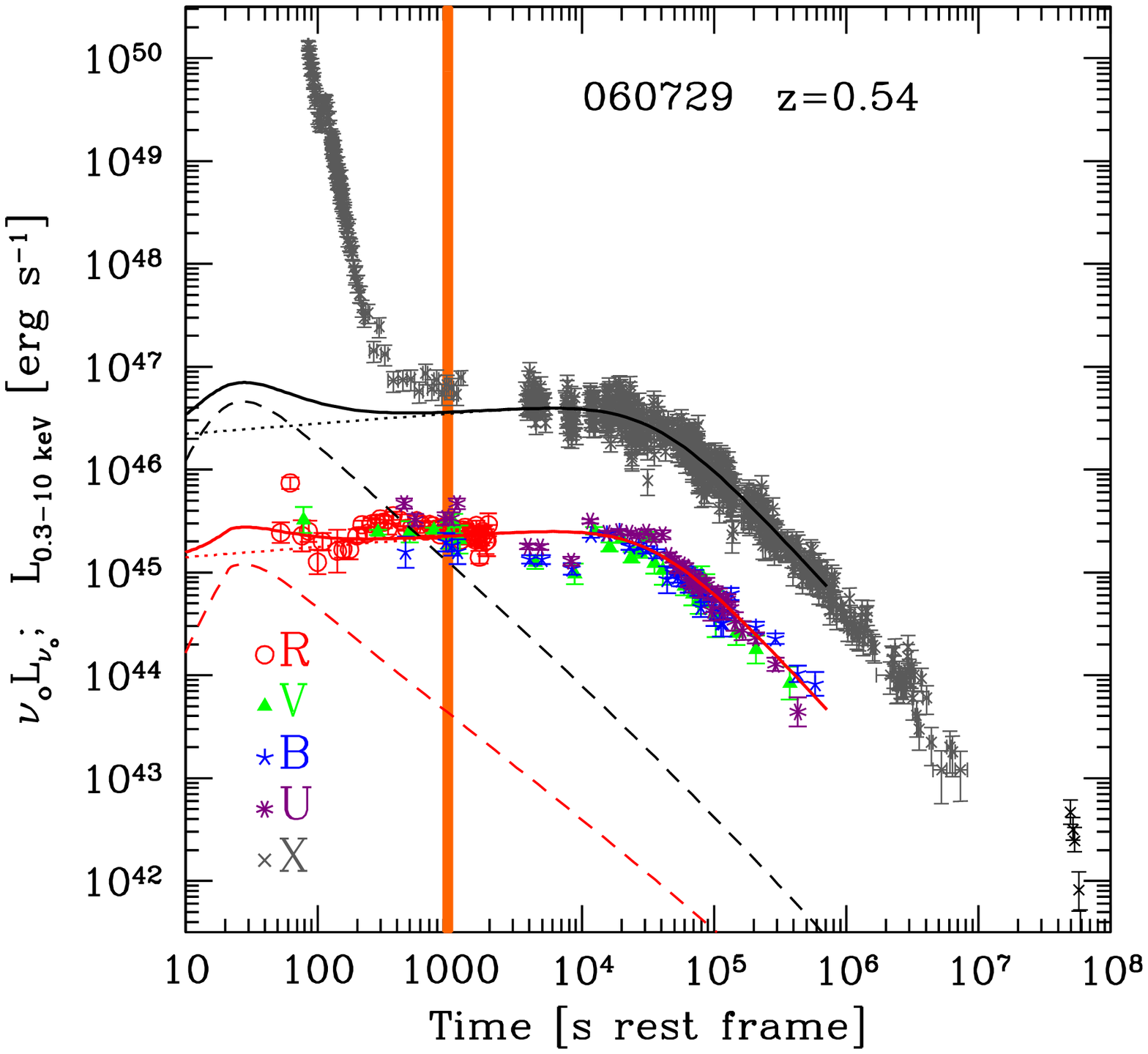,width=9cm,height=7.5cm} 
\vskip -0.5cm
\psfig{figure=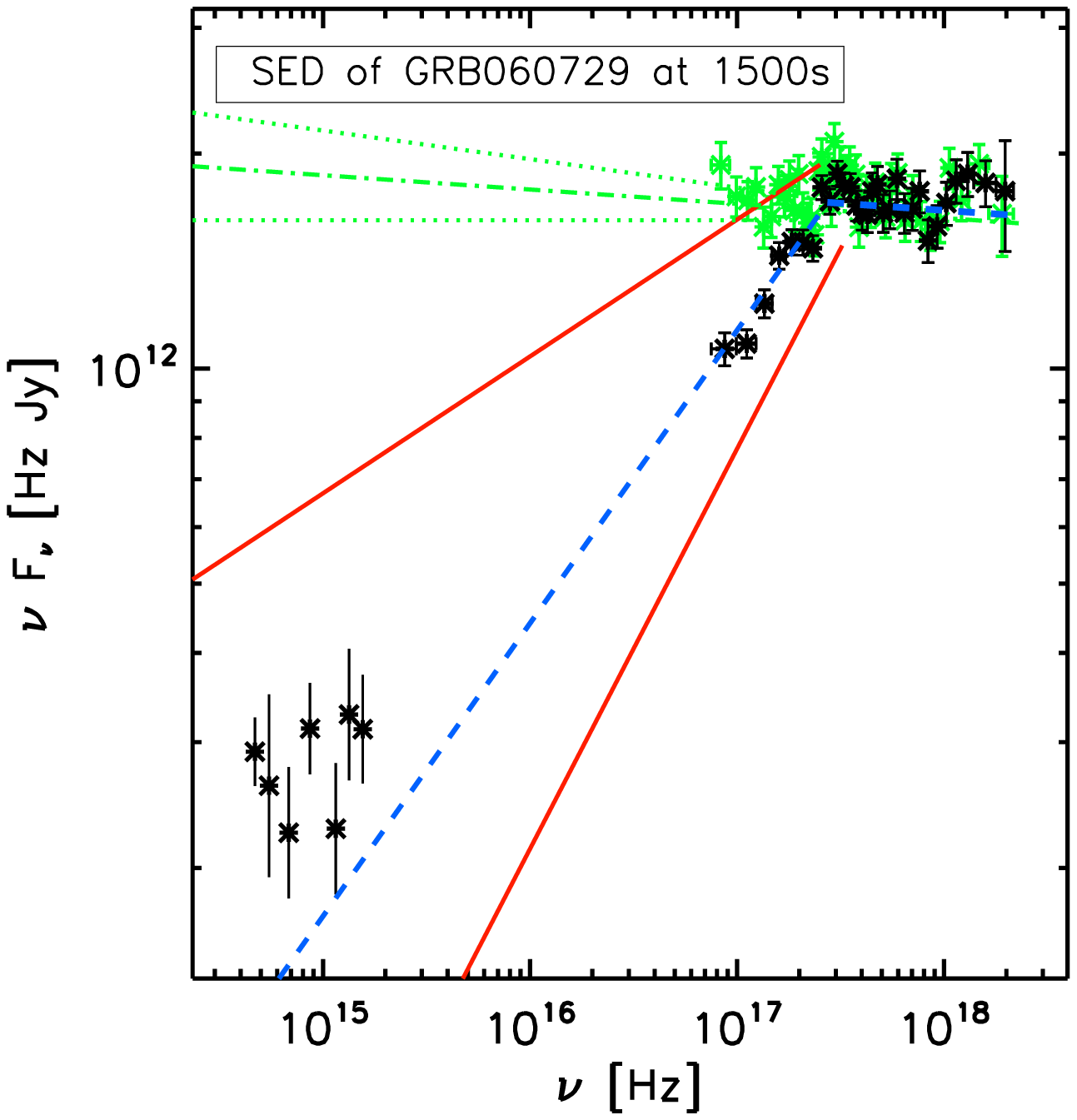,width=8.5cm,height=7cm}
\caption{
Top panel:
X--ray (in grey) and optical (different symbols, as labelled)
light curves of GRB 060729 in rest frame time. Same notation as in Fig. \ref{050802lc}.
 The vertical line represents the time at which the SED is extracted
Bottom panel:
Optical to X--ray $\nu F_{\nu}$ SED at about
1500 s after trigger in the observer frame (970 s rest frame). 
The  dashed and solid lines show respectively the best fit (with the ABP model) to the X--ray spectrum (the spectral parameters are reported in Tab. 2) and the uncertainties on the slope of the low energy spectral index 
$\beta_{\rm X, 1}$. The dotted line shows  the best fit (with the AP model) to the X--ray spectrum.
} 
\label{060729lc}
\end{figure}

 The UVOT data in 6 filters are from  Grupe et al. (2007) while the
ROSTE $R$ band photometry is from Rykoff et al. (2009). 

 After a steep decay in X--rays for 
about 400 s, the optical and X--rays light curves track each other and  are 
characterised by a long lasting ($\sim$ 50 ks) 
shallow decay phase. Following an achromatic break, a steeper 
flux decay phase is observed in X--rays up to about $10^7$ s after 
the burst (Grupe et al. 2009).
The light--curves are 
dominated by the ``late prompt" component with no significant 
evolution of the optical to X--ray flux ratio as can be seen in Fig. 
\ref{060729lc}. 
As a consequence, the optical to X--ray SED is not expected to evolve in time.  

The bottom panel of Fig. \ref{060729lc}  
shows that the optical  flux at about 1500 s is indeed consistent with the  extrapolation to 
the optical band of the broken power law X--ray spectrum. The SED does not require 
any additional host galaxy dust absorption  and the  poorly constrained optical spectral index is consistent within errors with $\beta_{\rm X, 1}$.  The quality of the optical--UV SED is not good enough to directly constrain the  $A_{\rm V}^{\rm host}$. 

It would be possible to consider SEDs at later times based on the UVOT data (the 
$R$ band photometry covers only the first XRT orbit).  However since neither the optical to X--ray flux ratio 
nor the colour significantly evolve we present here only the most  complete SED at 1500 s.  In this case the plotted X--ray spectrum is extracted from the first two PC mode orbits excluding the first 150 s of the first orbit in order to avoid the contribution of the steep decay phase.

\subsection{GRB 061007}

\begin{figure}
\vskip -0.5cm
\psfig{figure=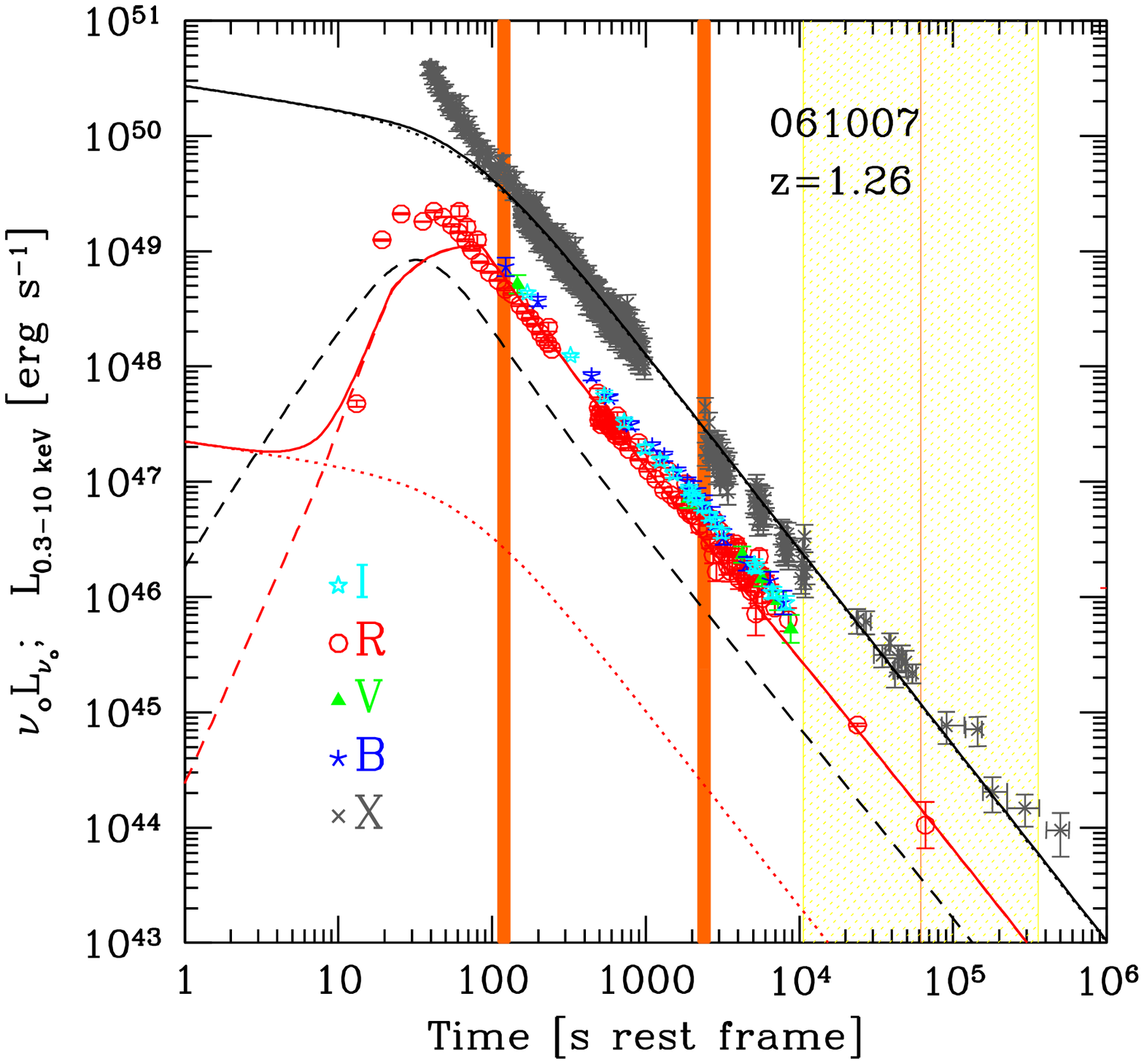,width=9cm,height=7.5cm} 
\vskip -0.5cm
\psfig{figure=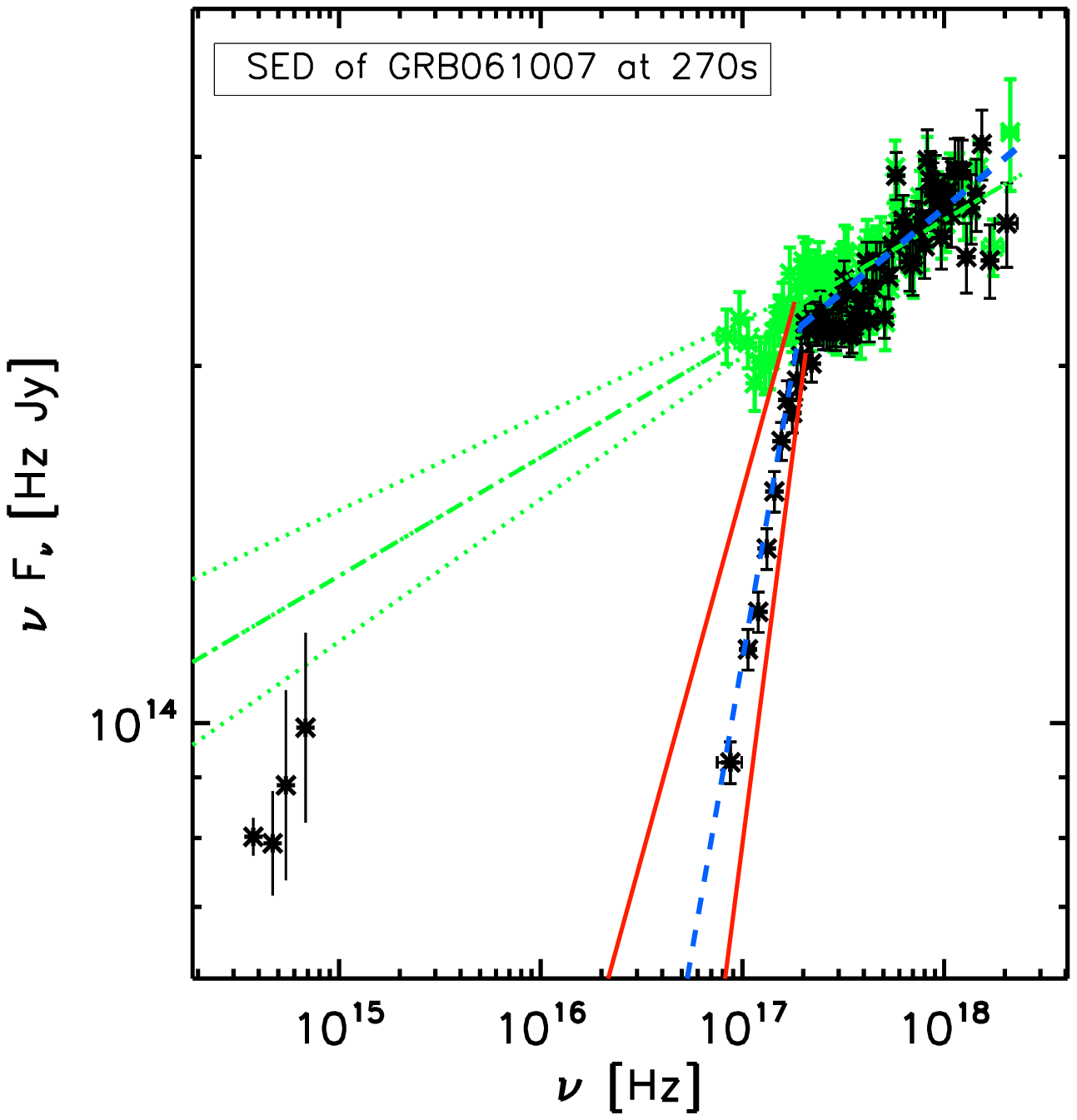,width=8.5cm,height=7cm}
\vskip -0.1cm
\psfig{figure=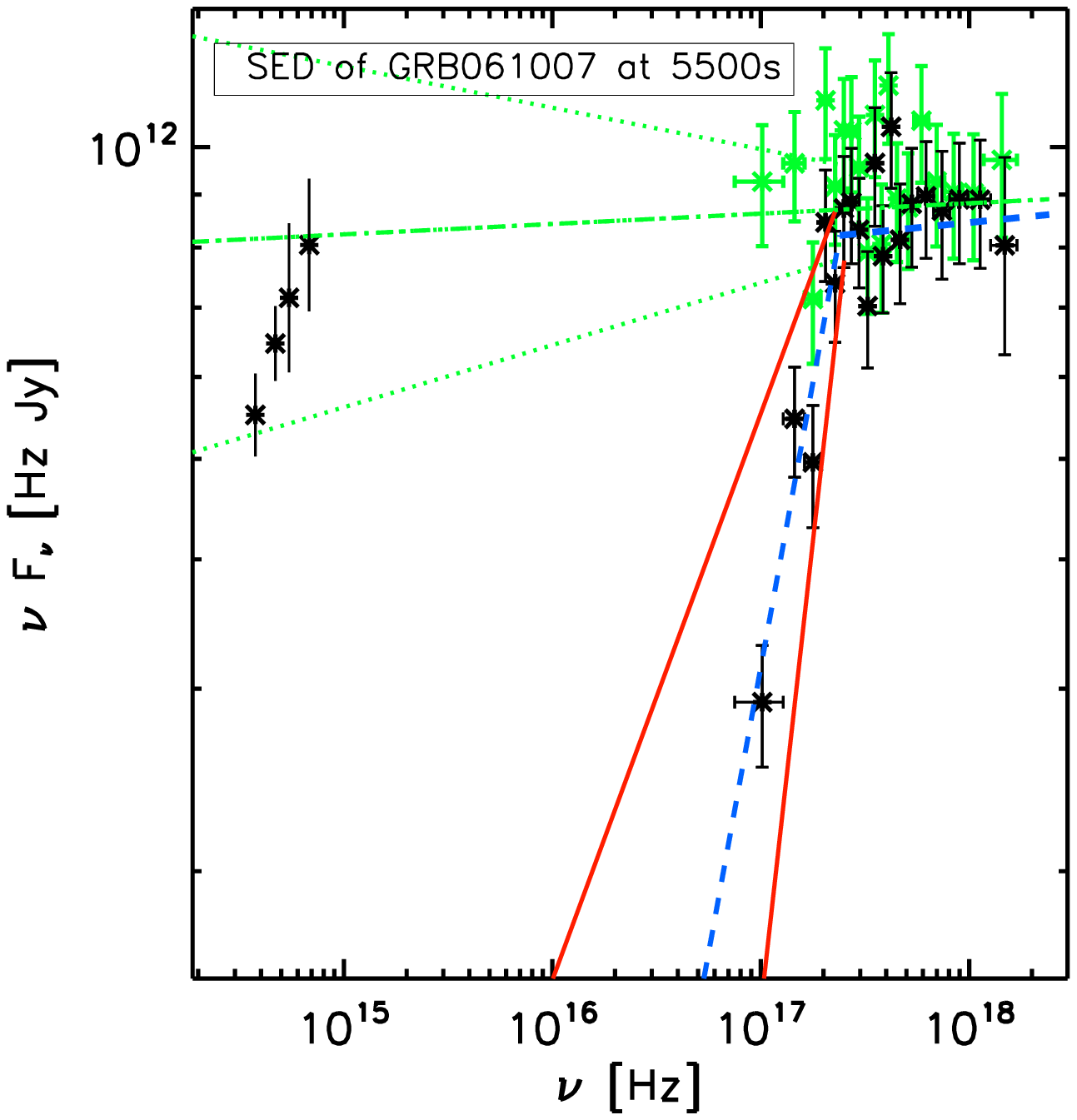,width=8.5cm,height=7cm}
\caption{
Top panel:
X--ray  and optical light curves of GRB 061007 in rest frame time. 
Same notation as in Fig. \ref{050802lc}.
 The vertical line and stripes indicate the jet times expected if the burst followed the 
$E_{\rm peak}$ vs. $E_{\gamma}$ ``Ghirlanda relation" (Ghirlanda et al. 2007) (see text).
 The vertical lines mark the times at which the SEDs are extracted.
Middle and bottom panels:
Optical to X--ray $\nu F_{\nu}$ SED at about 270 s (middle) and 5500 s (bottom)
after trigger in the observer frame (corresponding 
to 120 s and 2.4 ks in the rest frame). 
The  dashed and solid lines show respectively the best fit (with the ABP model) to the X--ray spectrum (the spectral parameters are reported in Tab. 2) and the uncertainties on the slope of the low energy spectral index 
$\beta_{\rm X, 1}$. The dotted line shows  the best fit (with the AP model) to the X--ray spectrum.
} 
\label{061007lc}
\end{figure}
 
The photometric data are from Mundell et al. (2007) 
($I$, $R$, $V$ and $B$ bands) and Rykoff et al. (2009) (ROTSE $R$ band). 
After a steeper  flux decay lasting 
about 90 s (rest frame), the X--ray light curve declines following a 
single  power--law for the whole observed time.  
The optical flux instead shows a fast rise ( by about 2 orders of magnitude) 
in the first 40 s followed by a simple power--law decay 
up to about 60 ks rest frame (see Fig. \ref{061007lc}). 
 The first $R$ band fluxes are simultaneous with the $\gamma$--ray prompt emission, and 
the rise between the first and second detection is faster than $t^5$,  
hardy explainable with any standard emission mechanism. 

 In the two component modelling  the optical light curve (after the end of the prompt 
phase, i.e. $\sim$ 50s in the rest frame) is dominated by the ``standard 
afterglow" emission.  
The single power--law X--ray decay phase after the end of the 
$\gamma$--ray detection would be  dominated by the ``late prompt" component. 
 Thus the X--ray and optical fluxes would be dominated by two different
components, despite the similarity of the light curves after $\sim$100 s (rest frame), 
requiring a hard  ``late prompt" $\beta_{\rm X,1}$ in order for this emission 
not to significantly contribute to the observed optical flux.

The optical  fluxes
have been corrected for  a host galaxy dust extinction $A_{\rm V}^{\rm host}=0.54 \pm 0.30$
(Kann et al. 2009). 
We considered two SEDs at the times where all of the four photometric bands are simultaneously available: 
the first one at about 270 s (observed frame), immediately after 
the beginning of the simple power--law X--ray decay, and the second 
one at about 5.5 ks (observed frame).  The X--ray data in the latter SED are extracted from the first two orbits in PC mode while the ones plotted in the first SED are extracted from the time interval reported in Tab. 1. 
 In both cases the hard $\beta_{\rm X, 1}$  found with the broken power law fitting (see Tab. 2) implies a negligible contribution of the X--ray component in the optical band,  supporting the proposed interpretation as can be seen from the middle and bottom panels of Fig. \ref{061007lc}.

The X--ray light curve  does not show any slope variation in correspondence to 
the expected jet break time obtained in the assumption that the GRB follows the ``Ghirlanda relation". 
Once again this is in agreement with the  ``late prompt" dominated nature of the
X--ray flux. The jet break should instead be visible in the optical, 
but unfortunately there are no observations after 150 ks (observer frame) to confirm or rule out this prediction. 

 The early time  optical to X--ray SEDs  of GRB 061007 has been  analysed also by Schady et al. (2007b) and by Mundell et al. (2007). They extracted the SED around 600 and 300 s observer frame after trigger respectively and they found these SEDs  to be well fitted by a single  power-law. As can be seen in fig. \ref{061007lc}, the optical fluxes at 270 s (observer frame) are consistent with an extrapolation of a single power law fit of the X--ray spectrum. Their single power law X--ray fits give results consistent with the ones presented in tab. \ref{nh_av_pow}. In this paper we consider also the broad band SED of GRB 061007 at later times. The bottom panel of fig. \ref{061007lc} shows that after 5 ks the optical fluxes are no more consistent with an extrapolation of the X--ray data single power law fit and the  two component scenario that we considered in the light curve modelling is in good agreement with both the early and late time SEDs.

\subsection{GRB 061126}

\begin{figure}
\vskip -0.5cm
\psfig{figure=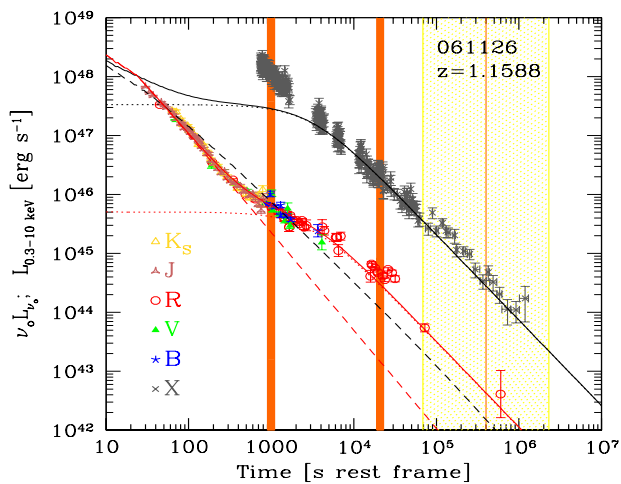,width=9cm,height=7.5cm} 
\vskip -0.5cm
\psfig{figure=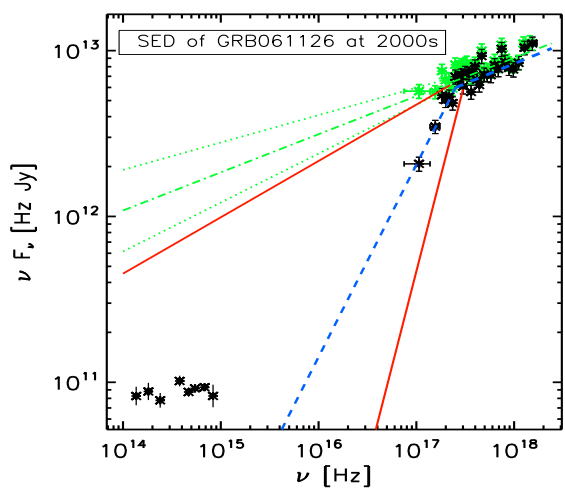,width=8.5cm,height=7cm}
\vskip -0.1cm
\psfig{figure=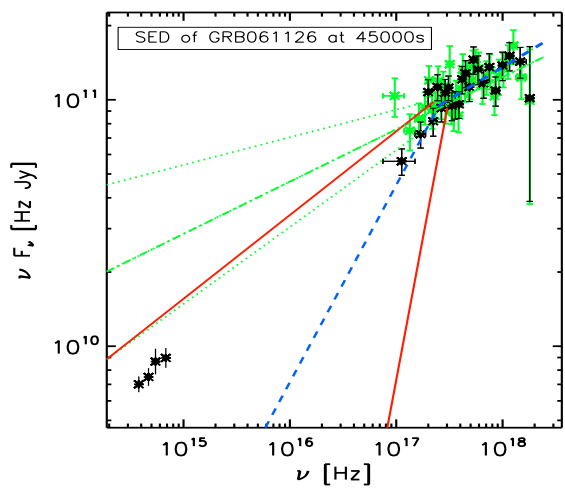,width=8.5cm,height=7cm}
\caption{
Top panel:
X--ray (in grey) and optical (different symbols, as labelled)
light curves of GRB 061126 in rest frame time. Same notation as in Fig. \ref{050802lc}.
The vertical line and stripes indicate the jet times expected if the burst followed the 
$E_{\rm peak}$ vs. $E_{\gamma}$ ``Ghirlanda relation" (Ghirlanda et al. 2007) (see text).
 The vertical lines mark the times at which the SEDs are extracted.
Middle and bottom panels:
Optical to X--ray $\nu F_{\nu}$ SED  at about 2200 s (middle) and 45 ks (bottom)
after trigger in the observer frame (corresponding to 1000 s and 21 ks in the rest frame). 
The  dashed and solid lines show respectively the best fit (with the ABP model) to the X--ray spectrum (the spectral parameters are reported in Tab. 2) and the uncertainties on the slope of the low energy spectral index 
$\beta_{\rm X, 1}$. The dotted line shows  the best fit (with the AP model) to the X--ray spectrum.
} 
\label{061126lc}
\end{figure}

A very rich photometric sampling is available for GRB 061126 
(Perley et al. 2008,  Gomboc et al. 2008). 
After a steeper decay the X--ray light curve 
follows a single power--law flux decay for the whole observed time. 
The IR--optical--UV light curve instead shows a more complex behaviour, as shown  
in Fig. \ref{061126lc}. 

We  modelled the  power--law decay of the X--ray light curve 
as a  ``late prompt" component. 
The optical bands is accounted for by a sudden transition from a 
``standard afterglow" dominated early time 
to a  ``late prompt" dominated late time behaviour. 
If correct, this scenario would imply a spectral
evolution from a two components to a single component optical--to--X--ray SED. 
We would also expect an evolution of $\beta_{\rm o}$ at the time of transition between 
the two components.

We extracted two SEDs, plotted in the middle and bottom panels of 
Fig. \ref{061126lc}. 
The first one (middle panel) corresponds to $\sim$2000 s (observer frame)
and is obtained using 8 contemporaneous photometric 
bands ($U$,  $B$, $V$, $R$, $I$, $J$, $H$, and $Ks$).  The X--ray data are extracted from the time interval reported in Tab. 1.   At this time the optical spectrum is well fitted by 
a single power law with $\beta_{\rm o}=0.94\pm 0.05$, and no host galaxy dust absorption is required.
This is in agreement with the findings by Perley et al. (2008). 
The optical spectrum slope is inconsistent with the X--ray spectrum. 
We examined a second SED at 45 ks (observer frame; bottom panel of Fig.
\ref{061126lc}) when both the optical 
and X--ray light curves are dominated by the ``late prompt" component. The plotted X--ray data are extracted from the time integrated spectrum of the second XRT observation.
At that time only 4 optical bands are available ($I_c$, $R_c$, $V$ and $B$), 
and the spectrum is still well fitted by a single power--law without 
host galaxy dust absorption, but the spectral index is harder 
than at earlier times (i.e. $\beta_{\rm o}=0.54\pm 0.1$).  
The bottom panel of Fig.  \ref{061126lc} reveals that not only the optical fluxes  
but also the optical slope are now consistent with the extrapolation from the X--rays 
(with slope $\beta_{X,1})$.

As predicted by the light curve modelling, the optical spectral index evolves after the
transition from standard afterglow to late prompt emission, with the SED becoming consistent 
with a single dominating component. 

We can contrast our  interpretation with the alternative one 
proposed by Gomboc et al. (2008), who suggested that 
the presence of some dust absorption at early times could account for the 
optical spectrum being consistent with a broken power--law optical--to-- X--ray SED. 
At later times the SED could be fitted with a similar 
broken power--law but without the need of any host galaxy dust absorption. 
Thus in the Gomboc et al. (2008) scenario a change in dust
absorption would be required to interpret 
the optical spectral and broad band SED 
evolution. 

Also for this burst we can evaluate the expected jet break time in the hypothesis that 
the GRB follows the ``Ghirlanda relation". However, 
at the corresponding epoch both the 
optical and X--ray emission are dominated by the late prompt contribution
and thus  no jet break would be observable. 
This is indeed in agreement with the absence of a break in the observed light curves,
although the observations end soon after the predicted jet break time.

\section{Conclusions}

We analysed the 
{\it Swift} XRT data of a sample of 33 long GRBs selected by G09 to have known redshift, published 
estimate of the host galaxy dust absorption and good XRT and optical follow up. 

If the XRT 0.3--10 keV  spectra are modelled  as a single power law, 
we confirm that the host frame $N_{\rm H}^{\rm host}$ 
column densities are rather large  when compared to the values
of the host galaxy dust absorption inferred from the optical analysis, according to ``standard" 
extinction laws (see also Schady et al. 2007, Stratta et al. 2004). 

 For the 15 brightest bursts we could model the X--ray data with 
a broken power law, and in 7 cases we find  evidence of a spectral break (90\% confidence level). 
In such cases the required $N_{\rm H}^{\rm host}$ is in turn smaller than 
for the single power--law fitting and is marginally consistent with the column estimated by the optical extinction.
 However, in other 8  bright GRBs the X--ray spectrum does not show  
any break and some of them do require a large value of $N_{\rm H}^{\rm host}$. 
Therefore the presence of an intrinsic curvature in the 
spectrum cannot be considered as a general solution for the ``excess" 
of $N_{\rm H}^{\rm host}$ commonly found in GRB X--ray spectral analysis.

 In order to test the interpretation by G09 that the X-ray and optical light curve complex 
behaviour can be 
interpreted as due to the contributions to the emission by two different components, 
we combine the results of the light curve de-convolution with the X--ray and broad band spectral properties
at different times.

 In particular we checked whether the presence of a break in the XRT spectra 
is consistent with what has been observed in the optical bands by 
studying the time dependent optical to X--rays SEDs
of the GRBs for which a spectral break was found. 

 We found that 7 of the 8 events are consistent with the presence of a break 
in the XRT spectra and the evolution of the broad band SEDs appears to support
the predictions of the two components scenario, even in the
presence of complex light curve behaviours. In one case (i.e. GRB 060210), the quality of the data does not allow us to solve the ambiguity between the temporal and spectral analysis.

 Consistency is also found in relation with the (lack of) evidence for 
jet breaks in the light 
curves, whose break time is estimated by assuming that the Ghirlanda relation holds for all GRBs (Ghirlanda et al. 2007). 
Indeed  light curves are observed to steepen in  
correspondence with the jet break time only when the light 
curve is dominated by the  ``standard afterglow" emission while no 
break is detected if the other component is dominating. 

 Further testing of the two--component modelling 
requires to extend the simultaneous multi--band optical 
follow--up at later times (i.e. several days after the trigger), 
when the typical expected $R$ band magnitudes are around 24--25. 
Such an intensive and long lasting multi--band follow up
could allow us: i)  to search for possible optical spectral 
index evolution when a different component becomes dominant 
(as in the case of GRB 061126); ii)  
to test for the presence (absence) of 
jet breaks in ``standard afterglow'' (``late prompt'') dominated 
optical light--curves in a larger sample of events, presumably shedding
light also on the jet geometry and energetics.
   
\section{Acknowledgements}
We would like to thank Fabrizio Tavecchio for useful discussions and the anonymous referee for his interesting comments and suggestions.  We acknowledge ASI grant I/088/06/0. 
MN and AC also acknowledge the MIUR for partial financial support. This work made use of data supplied by the UK Swift Science Data Centre at the University of Leicester.


\begin{thebibliography}{}
\bibitem[]{} Berger E., Gladders M., Oemler G., 2005a, GCN, 3201
\bibitem[]{} Berger E., Gladders M. \& Oemler G., 2005b, GCN, 3201
\bibitem[]{} Berger E. \& Gladders M., 2006, GCN, 5170
\bibitem[]{} Bloom J.S., Foley R.J., Koceveki D. \& Perley D., 2006, GCN, 5217
\bibitem[]{} Bloom J. S., Perley D., Chen H. W., 2006, GCN, 5825
\bibitem[]{} Bloom J. S. et al., 2009, ApJ., 691, 723 
\bibitem[]{} Burlon D. et al., 2009, ApJ, 685, L19
\bibitem[]{} Butler N.R. \& Kocevski D., 2007, ApJ, 663, 407
\bibitem[]{} Cenko S. B., Kulkarni S. R., Gal-Yam A., Berger E., 2005, GCN, 3542
\bibitem[]{} Cenko S. B. et al., 2006a, ApJ, 652, 490
\bibitem[]{} Cenko S. B., Berger E., Cohen J., 2006b, GCN, 4592
\bibitem[]{} Cenko S. B. et al., 2009, ApJ, 693, 1484 
\bibitem[]{} Chen H. W., Thompson I., Prochaska J. X., Bloom J., 2005, GCN, 3709
\bibitem[]{} Covino S. et al., 2008, MNRAS, 388, 347
\bibitem[]{} Cucchiara A., Fox D. B., Berger E., 2006, GCN, 4729
\bibitem[]{} Curran P. A. et al., 2007, A\&A, 467, 1049
\bibitem[]{} Curran P. A., Starling R.L.C., van der Horst A.J., \& Wijers R.A.M.J.,  2009, MNRAS,  395, 580
\bibitem[]{} De Pasquale M. et al., 2007, MNRAS, 377, 1638
\bibitem[]{} De Pasquale M. et al., 2009, MNRAS, 392, 153 
\bibitem[]{} Ellison S. L. et al., 2006, MNRAS, 372, L38
\bibitem[]{} Evans P.A., Osborne J.P., Burrows D.N., \& Barthelmy
  S.D., 2008, GCN, 7955
\bibitem[]{} Evans P.A., Beardmore A.P., Page K.L., et al., 2009,
  MNRAS submitted (arXiv:0812.3662v1)
\bibitem[]{} de Ugarte Postigo A. et al., 2007, A\&A, 462, L57  
\bibitem[]{} Dickey J.M., \& Lockman F.J., 1990, ARA\&A, 28, 215
\bibitem[]{} Foley R. J., Chen H. W., Bloom J., Prochaska J. X., 2005, GCN, 3483
\bibitem[]{} Fox D. B., Berger E., Price P. A., Cenko S. B., 2007, GCN, 6071
\bibitem[]{} Fugazza D., DÕAvanzo P., Malesani D., 2006, GCN, 5513
\bibitem[]{} Fynbo J.P.U., Hjorth, J., Jensen, B.L., Jakobsson P.,
             Moller P. \& N\"ar\"anen J., 2005a, GCN, 3136
\bibitem[]{} Fynbo J.P.U., Jensen, B.L.,  Hjorth, J. et al., 2005b, GCN, 3176
\bibitem[]{} Fynbo J. P. U. et al., 2005c, GCN, 3749
\bibitem[]{} Fynbo J. P. U. et al., 2005d, GCN, 3874
\bibitem[]{} Fynbo J. P. U. et al., 2005e, GCN, 3756
\bibitem[]{} Fynbo J.P.U., Limousin M., Castro Cer\'on J.M., Jensen B.L. \& 
             N\''ar\''anen J., 2006a, GCN, 4692
\bibitem[]{} Fynbo J. P. U. et al., 2006b, GCN, 5651
\bibitem[]{} Galama T. J., Wijers R. A. M. J., 2001, ApJ, 549, L209
\bibitem[]{} Gehrels N. et al., 2004, ApJ, 611, 1005
\bibitem[]{} Ghirlanda G., Nava L., Ghisellini G., Firmani C., 2007, A\&A, 466, 127
\bibitem[]{} Ghisellini G., Ghirlanda G., Nava L., \& Firmani C., 2007, ApJ, 658, 75
\bibitem[]{} Ghisellini G., Nardini M., Ghirlanda G., \& Celotti A., 2009, MNRAS 393, 253
\bibitem[]{} Godet O., Beardmore A.P., Abbey A.F., et al., 2009, A\&A,
  494, 775
\bibitem[]{} Gomboc et al., 2008, ApJ, 687, 443
\bibitem[]{}  Grupe D. et al., 2007, ApJ, 662, 443
\bibitem[]{} Holland S. T. et al., 2007, AJ, 133, 122
\bibitem[]{} Jakobsson P., et al., 2006, A\&A, 460, L13
\bibitem[]{} Jaunsen A. O., Malesani D., Fynbo J. P. U., Sollerman J., Vreeswijk P. M.,
2007, GCN, 6010
\bibitem[]{} Kalberla P.M.W., Burton W.B., Hartmann D., Arnal E.M.,
  Bajaja E., Morras R., \& P\"oppel W.G.L., 2005, A\&A, 440, 775
\bibitem[]{} Kann D.A., Klose S., \& Zeh A., 2006, ApJ, 641, 993
\bibitem[]{} Kann D.A., Klose S., Zhang B. et al., 2009, ApJ submitted
(2007arXiv0712.2186K)
\bibitem[]{} Mangano V. et al., 2007, A\&A, 470, 105
\bibitem[]{} Misra K., Bhattacharya D., Sahu D.K., Sagar R., Anupama
             G.C., Castro-Tirado A.J., Guziy S.S. \& Bhatt B.C, 2007,
             A\&A, 464, 903
\bibitem[]{} Moretti A., Campana S., Mineo T., et al., 2005, SPIE,
  5898, 360
\bibitem[]{} Mundell C. G. et al., 2007, ApJ, 660, 489
\bibitem[]{} Nardini M., Ghisellini G., Ghirlanda G., 2006, A\&A, 451, 821
\bibitem[]{} Nardini M., Ghisellini G., Ghirlanda G., 2008, MNRAS,
  386, L87
  \bibitem[]{} Nousek J. A. et al., 2006, ApJ, 642, 389
\bibitem[]{} Oates S. R. et al., 2007, MNRAS, 380, 270
\bibitem[]{} Osip D., Chen H. W., Prochaska J. X., 2006, GCN, 5715
\bibitem[]{} Page K. L. et al., 2007, ApJ, 663, 1125
\bibitem[]{} Panaitescu A., \& Kumar P., 2000, ApJ, 543, 66
\bibitem[]{} Pandey S. B. et al., 2006, A\&A, 460, 415
\bibitem[]{} Perley D. A., Chornock R., Bloom J. S., Fassnacht C., Auger M. W., 2007,
GCN, 6850
\bibitem[]{} Perley D. A. et al., 2008a, ApJ, 672, 449
\bibitem[]{} Perley D. A., Bloom J. S., 2008a, GCN, 7406
\bibitem[]{} Perley D. A., Bloom J. S., 2008b, GCN, 7535
\bibitem[]{} Perna R., Lazzati D., \& Fiore F., 2003, ApJ., 585, 775 
\bibitem[]{} Price P.A., Berger E. \& Fox D.B., 2006, GCN, 5275
\bibitem[]{} Prochaska J.X., Bloom J.S., Wright J.T., Butler R.P.,
             Chen H.W., Vogt S.S.  \& Marcy G.W., 2005, GCN, 3833
\bibitem[]{} Prochaska J.X., Chen H.W., Bloom J.S., Falco E., Dupree
A.K., 2006, GCN, 5002   
\bibitem[]{} Prochaska J. X., Perley D. A., Modjaz M., Bloom J. S., Poznanski D., Chen
H.-W., 2007, GCN, 6864          
\bibitem[]{} Prochaska J. X., Murphy M., Malec A. L., Miller K., 2008, GCN, 7388
\bibitem[]{} Protassov R., van Dyk D.A., Connors A., Kashyap V.L., \& Siemiginowska A., 2002, ApJ., 571, 545  
\bibitem[]{} Rol E., Jakobsson P., Tanvir N., Levan A., 2006, GCN, 5555
Romano P.
\bibitem[]{} Romano P., Campana S., Chincarini G., et al. 2006, A\&A,
  456, 917
\bibitem[]{} Ruiz-Velasco A. E. et al., 2007, ApJ, 669, 1
\bibitem[]{} Rykoff et al., 2009, ApJ submitted (	arXiv:0904.0261v1)
\bibitem[]{} Schady P., et al., 2007, MNRAS, 377, 273
\bibitem[]{} Schady P., et al., 2007b, MNRAS, 380, 1041
\bibitem[]{} Starling R.L.C., Vreeswijk P.M., Ellison S.L., 2005, A\&A, 442, L21
\bibitem[]{} Still M. et al., 2005, ApJ, 635, 1187
\bibitem[]{} Stratta G., Fiore F., Antonelli L. A., Piro L., \& De Pasquale M., 2004, ApJ, 608, 846
\bibitem[]{} Th\"one C.C., Levan A., Jakobsson P., et al., 2006 GCN, 5373
\bibitem[]{} Th\"one C. C. et al., 2008, A\&A, submitted (arXiv:0806.1182)
\bibitem[]{} Troja E. et al., 2007, ApJ, 665, L97
\bibitem[]{} Vaughan S. et al. 2006, ApJ, 638, 920
\bibitem[]{} Vreeswijk P. M., et al., 2007, A\&A, 468, 83
\bibitem[]{} Vreeswijk P. M., Smette A., Malesani D., Fynbo J. P. U., Jensen B. M.,
Jakobsson P., Jaunsen A. O., Ledoux C., 2008, GCN, 7444
\bibitem[]{} Watson D., Fynbo J. P. U., Ledoux C. et al., 2006, ApJ, 652, 1011
\bibitem[]{} Watson D., Hjorth J., Fynbo J.P.U., Jakobsson P.,
\bibitem[]{}  Sollerman J., \& Wijers R.A.M.J., 2007, ApJ., 660, 101 
\bibitem[]{} Zhang B. et al., 2006, ApJ, 642, 354
\bibitem[]{} Zhang B., 2007, Adv. Space Res., 40, 1186
\end{thebibliography}
\end{document}